\documentclass[aps,pra,twocolumn,amsmath,amssymb,showpacs,floatfix,superscriptaddress]{revtex4-1}

\usepackage{times}
\usepackage[utf8x]{inputenc}
\usepackage[english]{babel}
\usepackage[T1]{fontenc}
\usepackage{lmodern}
\usepackage{amssymb}
\usepackage{amsmath}
\usepackage{amsthm}
\usepackage[pdftex]{graphicx}
\usepackage{epstopdf}
\usepackage{braket}
\usepackage[bottom]{footmisc}
\usepackage{dcolumn}
\usepackage{bm}
\usepackage{color}
\usepackage{relsize,dsfont,mathrsfs,empheq,verbatim,upgreek}
\usepackage{etoolbox}
\usepackage[caption = false]{subfig}

\usepackage[hidelinks]{hyperref}
\usepackage{bbm}

\usepackage{enumerate}
\usepackage{enumitem}
\newcommand{\Tr}{\mathrm{Tr}}

\newcommand{\atanh}{\mathrm{atanh}}

\newcommand{\gen}{\mathcal{L}_t}

\begin{document}


\title{Local approach to entropy production in the nonequilibrium dynamics of open quantum systems}


\author{Irene Ada Picatoste}
\email{irene.ada.picatoste@physik.uni-freiburg.de}
\affiliation{Institute of Physics, University of Freiburg, Hermann-Herder-Stra{\ss}e 3, D-79104 Freiburg, Germany}

\author{Alessandra Colla}  
\affiliation{Institute of Physics, University of Freiburg, Hermann-Herder-Stra{\ss}e 3, D-79104 Freiburg, Germany}

\author{Heinz-Peter Breuer}
\email{breuer@physik.uni-freiburg.de}
\affiliation{Institute of Physics, University of Freiburg, Hermann-Herder-Stra{\ss}e 3, D-79104 Freiburg, Germany}
\affiliation{EUCOR Centre for Quantum Science and Quantum Computing, University of Freiburg, Hermann-Herder-Str. 3, D-79104 Freiburg, Germany}

\begin{abstract}
We discuss fundamental features of the local
expression for the entropy production rate of
the nonequilibrium quantum dynamics of open systems
and its relations to memory effects and the spectrum of the 
generator of the dynamics.
Defining the entropy production rate as negative rate of 
change of the relative entropy with respect 
to an instantaneous fixed point, it is shown 
that positivity of the entropy production 
rate for all possible initial states implies that the real parts of the 
eigenvalues of the time-local generator 
for the quantum master equation are always 
negative. It is further demonstrated that 
Markovian dynamics, identified as
P-divisibility of the quantum dynamical 
map, implies positivity of entropy production 
rate, thus providing a kind of generalized 
second law in the nonequilibrium regime. 
We also prove by means of the 
counterexample of a phase covariant quantum 
master equation that the converse of 
this statement is not true, i.e., there 
are non-Markovian dynamics for which 
the entropy production rate is always 
positive. Thus, we conclude that the 
emergence of negative entropy production 
rates is a sufficient but not necessary 
condition for non-Markovianity of the 
quantum dynamics. Finally, we also consider 
a recently introduced map-based notion of entropy production and show the equivalence between its positivity and Markovianity for general finite-dimensional systems.
\end{abstract}

\date{\today}

\maketitle


\section{Introduction}
\label{sec:intro}

The thermodynamics of strongly 
coupled open systems constitute a 
new and exciting field of research, 
with many foundational questions yet 
to be answered \cite{BinderBook,StrasbergBook,GemmerBook}. 
In the literature
plenty of different proposals exist to address the problems of 
defining an energy, heat, work, 
and entropy production in processes 
governed by non-unitary dynamics 
\cite{Esposito2010Jan,Colla2022May,
Nicacio2023Aug, Elouard2023_QT}, but 
a general consensus has still to 
emerge. 
 
The entropy of a system 
measures its degree of disorder, but 
also represents the available 
information about the system, or the knowledge 
we have of it.  
In closed systems following unitary 
evolution this is a conserved quantity, but 
when the system comes into contact with 
an external environment, the reduced 
description of the system or of the environment 
no longer contains the 
full information. Specifically,
the information about the correlations 
built between the both is lost, and 
the entropy of the system can change in time.
Typically, some of this change can be ascribed to 
an entropy flow coming from the interaction 
with the environment, and it is the mismatch 
between these two quantities that we call  entropy production \cite{Altaner2017Oct}, 
which is further
related to 
the degree of irreversibility of the process.

Despite significant progress in quantum thermodynamics, a universally accepted definition of entropy production in the strong-coupling regime is still lacking. Several proposals exist in the literature \cite{Esposito2010Jan, Elouard2023_QT, Rivas2020Apr}, but they do not establish a clear link between memory effects in the dynamics and violations of the second law \cite{Strasberg2019Jan}. In this work, we employ a generalized entropy production rate and demonstrate that it provides exactly such a connection: Markovianity guarantees positivity
of the entropy production rate, giving us 
a generalized second law for the quantum regime.
We also study the converse and prove that 
Markovianity is a sufficient, but not necessary 
condition for a positive entropy production rate.
We show this by means of an example phase-covariant dynamics of a qubit, for which 
Markovianity is violated while the entropy 
production rate for all initial states remains always positive. 
This framework clarifies the interplay between irreversibility and memory effects, and allows for a precise characterization of the thermodynamic behavior of open quantum systems beyond the weak-coupling approximation. 

Our adopted generalized  
entropy production rate  extends 
the previously established weak 
coupling formulation \cite{Spohn1978Jan}, 
and is given by the
negative rate of 
change of the
relative entropy between the state of the 
system and an instantaneous fixed point. 
Furthermore, one of the advantages of this definition is 
that it is a local expression, based on only 
the degrees of freedom of the system, and 
thus experimentally accessible. 
This formulation 
has already been proposed in some 
specific scenarios, e.g. for coarse-grained
processes \cite{Strasberg2019Jan}, 
pure decoherence dynamics \cite{Picatoste2025_decoherence}
or a Gaussian bosonic model under the
rotating-wave-approximation  
\cite{Colla2024dec}. In the present work, 
we take it as a general definition 
of entropy production. 

We further study  
the connection between entropy production 
rate and the spectrum of the generator of 
the dynamics, and show that negativity of 
the real part of the eigenvalues of the 
generator is a necessary condition for a 
positive entropy production rate at all times
for all initial states. 

Additionally, we look at a recent formulation
that extends the definition of entropy production such that
it becomes a property of the dynamical map \cite{Theret2026}. 
This formulation has so far been studied for phase-covariant 
qubit dynamics, and here we study its connection to Markovianity 
for general $d$-level systems. 

The article is structured as follows: 
in Sec. \ref{sec:open_quantum_systems} we 
introduce the general framework for the 
study of open quantum systems, and 
review the 
theory underlying  
dynamical maps and time-local 
generators (Sec. \ref{sec:dyn_maps_and_gen}), 
the spectral properties of the generator 
(Sec. \ref{sec:spectral_properties}) and 
non-Markovianity in the quantum regime 
(Sec. \ref{sec:non_markovianity}). 
Next, in Sec. \ref{sec:EPR_semigroups} we review 
the established definition of entropy 
production for quantum dynamical semigroups, 
which is 
followed by Sec. \ref{sec:EPR_generalized} 
where we generalized to strong-coupling 
scenarios. In the following 
Sec. \ref{sec:neg_eigvals} we show that, 
with this generalized definition, negativity 
of the real part of the eigenvalues of the 
generator is a necessary condition for an 
always 
positive entropy production rate.
In Sec. \ref{sec:EPR_and_Pdiv} we study 
the relation between entropy production 
and memory effects. Starting in Sec. 
\ref{sec:Pdiv_means_second_law} we prove that 
Markovianity is a sufficient condition for a 
positive entropy production rate, whereas in 
Sec. \ref{sec:secondlaw_doesntmean_Pdiv} we 
prove that it is not necessary by means of a 
counterexample. In Sec. \ref{sec:Camille} we 
take a look at the recent extension of the concept of 
entropy production to the dynamical map. 
Finally, we conclude in Sec. 
\ref{sec:conclusions}.


\section{Open quantum systems}
\label{sec:open_quantum_systems}

We consider a global, closed bipartite system 
with density matrix $\rho_{SE}(t)$ composed 
of an open system $S$ and an environment $E$ 
\cite{Breuer2002}. The 
composition thus evolves according to von Neumann's 
equation which reads (setting $\hbar = 1$)
\begin{eqnarray}\label{eq:von_Neumann}
    \frac{d}{dt} \rho_{SE}(t) = - i 
    \left[ H(t), \rho_{SE}(t) \right],
\end{eqnarray}
with the microscopic global Hamiltonian consisting of
that of the bare system and environment,
$H_S(t)$ and $H_E$, and the interction
Hamiltonian $H_I(t)$:
\begin{eqnarray}\label{eq:microscopic_Hamiltonian}
    H(t) = H_S(t) + H_E + H_I(t).
\end{eqnarray}

\subsection{Dynamical maps and time-local 
generators}
\label{sec:dyn_maps_and_gen}

The open system can be described through the 
reduced density matrix $\rho_S(t) \in \mathcal{S}_S$,
where $\mathcal{S}_S$ denotes the set of system states
and the reduced state is obtained 
by tracing over the environmental degrees of 
freedom, $\rho_S(t) = \Tr_E \left\{ \rho_{SE}(t) \right\}$.
The evolution of the system is then given 
by the dynamical map $\Phi_t$, which 
propagates the initial state to 
its future state:
\begin{eqnarray}\label{eq:dyn_map}
    \rho_S(t) = \Phi_t \left[ \rho_S(0) \right].
\end{eqnarray}
This map is always Hermiticity and 
trace preserving and, when the global initial state 
is uncorrelated $\rho_{SE}(0) = 
\rho_S(0) \otimes \rho_E(0)$,
then the dynamical map is additionally 
completely positive (CP).
There are many different 
ways to obtain 
the dynamical map: be it analytically 
when the full evolution is known 
\cite{Picatoste2025_decoherence,Smirne2010Aug},
through numerical techniques (see, for example,
\cite{Tanimura_HEOM,Tanimura2020,Suess_Strunz_HOPS}),
or experimentally by reconstructing 
the state of the system at each point in 
time \cite{Mohseni2008}. 

The map $\Phi_t$ thus takes an
initial state $\rho_S(0)$ and returns 
the evolved state $\rho_S(t)$. 
Assuming the map is invertible (which is 
generally the case, except perhaps at isolated 
points in time 
\cite{Breuer2002,Breuer2012Jul,Jagadish2023Oct}) we can further define 
a map
\begin{eqnarray}\label{eq:map_propagator}
    \Phi_{t,s} =
    \Phi_{t} \Phi_s^{-1}
\end{eqnarray}
which propagates a state from time $s$ to time $t$.
Since the inverse of the dynamical map is 
generally not CP, or even positive (P),
this propagator is also 
not CP in general, and also not P. 

The dynamics of the system are also 
governed by a first order linear differential 
equation, called a master equation, which 
can be obtained directly from the dynamical 
map by computing 
\begin{eqnarray}\label{eq:master_eq}
    \frac{d}{dt} \rho_S(t) = 
    \dot{\Phi}_t \Phi_t^{-1} \left[\rho_S(t)
    \right]
    \equiv \mathcal{L}_t \left[ \rho_S(t) \right].
\end{eqnarray}
The superoperator $\mathcal{L}_t$ is often called
time-convolutionless (TCL) or time-local 
generator of the master equation, and it 
typically exists except for isolated points 
in time when the dynamical map is not invertible.
It should be noted that this time-local master equation
provides an exact representation of the dynamics including
all effects from strong system-environment couplings
and non-Markovianity \cite{Breuer2016Apr}.


\subsection{Spectral properties of the 
generator}
\label{sec:spectral_properties}

Following \cite{Minganti2018}, we look into 
the spectral decomposition of the generator. 
Since we are concerned with time-dependent generators,
the eigensystem will also 
be time-dependent. At each point in time we 
have thus eigenvalues $\lambda_i(t)$ and 
eigenmatrices $\sigma_i(t)$ such that
\begin{equation}
    \mathcal{L}_t \left[\sigma_i(t) \right] 
    = \lambda_i(t) \sigma_i(t),
\end{equation}
with the set  
$\{\sigma_i(t) \}$ forming a complete basis.

The generator is not Hermitian, and 
consequently the 
eigenvalues can be complex and the 
eigenmatrices non-Hermitian. 
The generator is however Hermiticity 
preserving, which means that
\begin{equation}
    \left(\mathcal{L}_t \left[A \right]\right)^\dagger 
    = \gen \left[A^\dagger \right].
\end{equation}
This directly implies 
\begin{equation}
    \gen \left[\sigma_i^\dagger (t)\right] 
    = \lambda_i^* (t)\sigma_i^\dagger(t),
\end{equation}
which tells us that, if $\lambda_i(t)$ is real 
then $\sigma_i(t)$ can be chosen to be Hermitian and 
vice versa. Conversely, if $\lambda_i(t)$ 
is complex with eigenmatrix $\sigma_i(t)$, 
then $\sigma_i^\dagger(t)$ will also be an 
eigenmatrix with eigenvalue $\lambda_i^*(t)$.

The generator is also trace destroying, 
meaning
\begin{eqnarray}
    \Tr \{ \mathcal{L}_t [X] \} = 0 
    \hspace{0.5cm} \forall X.
\end{eqnarray}
Therefore, all eigenmatrices with 
non-zero eigenvalue
have zero trace.

An instantaneous fixed point (IFP) will be 
the eigenmatrix with eigenvalue zero, 
which we denote by $\rho_S^\star(t)$:
\begin{eqnarray}
    \mathcal{L}_t [\rho_S^\star(t)] = 0.
\end{eqnarray}
An IFP always exists, which can be proven by the 
following argument: 
We can define the adjoint generator as
the superoperator with the property
\cite{Breuer2002}, 
\begin{eqnarray} 
    \Tr \{ A \mathcal{L}_t [\rho_S] \} = 
    \Tr \{ \mathcal{L}_t^\dagger [A] \rho_S\},
\end{eqnarray}
for all system operators $A$ and density 
matrices $\rho_S$. 
Taking $A = \mathbbm{1}$ we see that 
$\mathcal{L}_t^\dagger [\mathbbm{1}] = 0$ (because
the generator is trace destroying),
that is, the identity is always an eigenoperator 
of the adjoint generator with eigenvalue $0$. 
It follows that there always
exists a Hermitian eigenoperator of the 
generator with eigenvalue zero, that is, 
an instantaneous fixed point \cite{Minganti_2019}.

This matrix has non-zero trace, for 
completeness of the basis, and 
the normalization can be chosen 
such that the eigenoperator has trace 1. 
Still, this doesn't ensure that the IFP is 
a state, since we are missing the requirement of 
positivity. We leave this as a subject of 
further research and treat in the following 
only the case for which the 
IFP is a state, and 
further assume that it is full-rank 
(i.e., it is not on the boundary 
of the state space). 


\subsection{Non-Markovianity}
\label{sec:non_markovianity}

There are many proposals in the literature as to how to define non-Markovian
dynamics in the quantum regime \cite{Rivas_2014, Breuer2016Apr, LI2018}. Here, we follow 
the definition based on the flow of information between the open system and its environment 
\cite{Breuer2009,Laine2010}. This information flow is defined in terms of the
distinguishability of two quantum states $\rho_S^{1,2}(t)$ quantified be the trace distance
\cite{Nielsen2000,Hayashi2006}
\begin{eqnarray}
    D_{\text{tr}} (\rho_S^1(t), \rho_S^2(t)) 
    = \frac{1}{2} \text{Tr} \left|\rho_S^1(t) - \rho_S^2(t) \right|.
\end{eqnarray}
Any dynamical decrease of the trace distance can be interpreted as a loss
of information in the open system, while a temporal increase of the trace distance
corresponds to a backflow of information from the environment to the open system
\cite{Breuer2009}. Consequently, a given dynamics is said to be Markovian if the
trace distance decreases monotonically, corresponding to a continuous loss of information 
from the open system to the environment. Thus, non-Markovian dynamics is characterized by
a non-monotonic behavior of the trace distance for some
pair of initial states $\rho_S^{1,2}(0)$. This concept also leads to a measure for the
degree of memory effects defined by
\begin{eqnarray}\label{eq:nonmarkov_tracedist}
    \mathcal{N}_\text{tr} (\Phi) = \max_{\rho_{S}^{1,2}(0)} 
    \int_{\dot{D}_\text{tr} > 0} dt \; \dot{D}_\text{tr}(t)
\end{eqnarray}
with $\dot{D}_\text{tr} (t) = \frac{d}{dt} D_\text{tr}
(\Phi_t [\rho_{S}^1(0)], \Phi_t[\rho_{S}^2(0)])$, measuring the total backflow
of information maximized over all pairs of initial states \cite{Laine2010}.

The trace distance provides a measure for the distinguishability of two quantum states
$\rho_S^1$ and $\rho_S^2$ given with equal a priori probabilities $p_1=p_2=\frac{1}{2}$.
One can introduce a certain bias such that $\rho_S^1$ is prepared with probability $p_1$, 
and the other state with probability $p_2=1-p_1$. In this case the distinguishability of the
states is quantified by the trace norm $||\Delta||=\Tr |\Delta|$ of the matrix
\begin{eqnarray}
    \Delta \equiv p_1 \rho_S^1 - p_2 \rho_S^2
\end{eqnarray}
which is known as Helstrom matrix \cite{Helstrom1976, Helstrom1967a}.
More precisely, this statement means that the maximal probability for a successful
state discrimination is given by the expression 
$p_{\text{max}} = \frac{1}{2}(1+||\Delta||)$. To include such biased cases one
replaces the trace distance of two quantum states by the trace norm of the corresponding
Helstrom matrix in the definitions above, where the definition of the measure according 
to Eq.~\eqref{eq:nonmarkov_tracedist} additionally involves a maximization over the
probability distribution $p_{1,2}$.

Within the present article we are using this definition of non-Markovianity based on the
Helstrom matrix. As shown in \cite{Wissmann2015,Chruscinski2011a} it
is equivalent to the P-divisibility of the dynamical map $\Phi_t$. We recall that
a dynamical map $\Phi_t$ is said to be P-divisible if the map $\Phi_{t,s}$ defined in 
Eq. \eqref{eq:map_propagator} is positive 
for all $t \geq s \geq 0$, where we assume that the inverse $\Phi_t^{-1}$ of the map 
always exists. Thus, in the following we will identify non-Markovianity with 
violation of P-divisibility of the map, and any 
P-divisible (invertible) evolution will be called Markovian.


\section{Entropy production  
for open quantum systems}
\label{sec:EPR}

Having established the framework for the
study of open quantum systems, 
we are now interested in a formulation of 
the entropy production. 
We begin by recalling the established formulation 
for semigroups (e.g. when the Born-Markov 
approximation is valid), and then 
extend this definition for systems 
outside of the weak-coupling regime. 

\subsection{Entropy production for 
dynamical semigroups}
\label{sec:EPR_semigroups}

We will base our definition of entropy 
production on its weak-coupling formulation 
developed by Spohn \cite{Spohn1978May,Spohn1978Jan}.
This approach is valid for CP semigroups, i.e. for dynamical maps with the 
property that $\Phi_{t+s} = \Phi_t \Phi_s$.
This implies that the generator is 
time-independent (and 
thus, also any fixed point) \cite{Breuer2002}. 
Additionally, if  
there is a unique fixed point, 
any initial state will converge to it in the long time limit, 
i.e. $\lim_{t\rightarrow \infty}\rho_S(t) = 
\rho_S^\star$.
Then, entropy production rate relative 
to a fixed point $\rho_S^\star$ and for an initial 
state $\rho_S(0)$
is defined as 
\begin{eqnarray}\label{eq:Spohn_EPR}
    \sigma_S(t) =- \left. \frac{d}{d\tau} 
    \right|_{\tau=0}
    D\left(\Phi_\tau [\rho_S(t)] || \rho^\star_S
    \right),
\end{eqnarray}
where 
\begin{eqnarray}
    D(\rho_1 || \rho_2) = \Tr \{\rho_1 \left(
\log \rho_1 - \log \rho_2 
\right)\}
\end{eqnarray}
is the relative entropy. 
The relative entropy
is also an expression 
which measures how difficult it is to 
distinguish two quantum states 
\cite{Vedral2002_relativeentropy_review}, such that
Eq. \eqref{eq:Spohn_EPR} tells us 
that entropy production rate is the negative rate of 
change of the distinguishability (measured by 
the relative entropy) between 
any state and the fixed point.

Because every CP semigroup is CP-divisible, 
and relative entropy is a contraction 
under CP maps, 
it can be shown that entropy 
production rate is always positive 
\cite{Spohn1978Jan}. 


\subsection{Generalized expression for entropy 
production}
\label{sec:EPR_generalized}

We now want to find an expression for 
entropy production rate for scenarios 
where the master equation is allowed to 
become explicitly time-dependent. This allows the
study of strong-coupling scenarios, non-Markovian 
dynamics or driven systems, among others. 

We will make use of two assumptions: that there 
is a unique fixed point at each point in time, 
and that the fixed point is always a state.
The first could be relaxed if we define, 
like Spohn, the entropy production ``relative
to a certain fixed point'' \cite{Spohn1978May}.
The second we find a more limiting assumption, 
and more work should be carried out in 
this direction in order to 
overcome this limitation.

Then, we extend the formulation of Spohn 
to define entropy production rate as
\begin{eqnarray}\label{eq:EPR_Spohn_generalized}
    \sigma_S (t) &=& - \left. \frac{d}{d \tau} \right|_{\tau = 0}
    D(\Phi_\tau^{(t)}[\rho_S(t)] || \rho_S^\star (t)) 
    \nonumber \\
    &=& - \left. \frac{d}{d \tau} \right|_{\tau = 0}
    D(\rho_S(t+\tau) || \rho_S^\star (t)),
\end{eqnarray}
where we have made use of 
the instantaneous map 
\begin{eqnarray}\label{eq:instantaneous_map}
    \Phi_\tau^{(t)} \equiv e^{\tau \mathcal{L}_t}.
\end{eqnarray}
This expression considers the rate of 
approach, measured by the relative entropy, 
between the state of the system and the 
IFP at each point in time. 
If the system is 
approaching the fixed point (as measured 
by the relative entropy), we will find a 
positive entropy production rate, while, 
if it is instantaneously distancing itself, 
the entropy production rate will become negative.

We note that Eq. \eqref{eq:EPR_Spohn_generalized} can also 
be written as 
\begin{eqnarray}\label{eq:EPR_generator}
    \sigma_S(t) = - \Tr \left\{\mathcal{L}_t [\rho_S(t)]
    \left(\log \rho_S(t) 
    - \log \rho_S^\star (t)\right)\right\},
\end{eqnarray}
which shows explicitly that the entropy production rate is 
a functional of the state of the open system: $\sigma_S(t) = \sigma_S
[\rho_S(t),t]$. This definition has already been introduced 
in different contexts, for example in the work \cite{Theret2026}
which will be discussed in more detail in Sec.~\ref{sec:Camille}.
In many cases the definition of entropy production rate
is connected with the restriction 
that the IFP is a Gibbs state of some kind of Hamiltonian, 
be it the bare Hamiltonian of the system 
\cite{Deffner2011,Colla2022May}, 
the Hamiltonian of mean force \cite{Strasberg2019Jan}, 
or an effective renormalized Hamiltonian \cite{Colla2024dec}. 
This stems from the following: 
Eq. \eqref{eq:EPR_Spohn_generalized}
can be rewritten as 
\begin{eqnarray}
\label{eq:EPR_kinda_Clausius}
    \sigma_S(t) = \dot{S}_S(t) + 
    \Tr \{ \dot{\rho}_S(t) \log \rho_S^\star(t) \},
\end{eqnarray}
where $S_S(t) = - \Tr \{\rho(t) \log \rho(t) \}$ is 
the von Neumann entropy of the system at 
each point in time. 
With this formula 
we can interpret the entropy production 
rate as the addition of two quantities: 
the change of the entropy of the system 
$\dot{S}_S(t)$ plus an entropy flow 
$\Tr \{ \dot{\rho}_S(t) \log(\rho_S^\star(t) )\}$.
When the IFP can be written as a Gibbs state of 
some Hamiltonian with a certain temperature 
$\rho_S^\star (t) = e^{-\beta_x (t) H_x(t)} / Z_x(t)$, 
and this Hamiltonian is the 
same one used to calculate first-law quantities 
(namely, that heat is given by 
$\dot{Q}_S(t) = \Tr \{H_x(t) \dot{\rho}_S(t)\}$), 
then 
Eq. \eqref{eq:EPR_kinda_Clausius} takes the form 
of the Clausius entropy production rate: 
\begin{eqnarray}
    \sigma_S(t) = \dot{S}_S(t) 
    - \beta_x (t) \dot{Q}_S(t).
\end{eqnarray} 

Clausius is however just a special case of 
the formulation in Eq. \eqref{eq:EPR_Spohn_generalized}, 
and does not generally hold. 
One may ask then why not use Clausius
to formulate entropy production rate, and
we find various
reasons for this. For one, there are 
still many contesting definitions for 
the heat exchange in 
open quantum systems 
\cite{Esposito2010Jan,Nicacio2023_gauge_QT, 
Elouard2023_QT}, but the definition 
we propose is independent of our formulation 
for the first law. Furthermore, it is not clear how 
to define the temperature of the environment when it 
is evolving and strongly coupled to the system,
and some of the approaches proposed to address 
this problem require the calculation of the 
energy \cite{Strasberg2021_Clausiusfinite} 
or entropy \cite{Elouard2023_QT}
of the bath, which may be convoluted or outright 
impossible in an experimental setup when the bath 
cannot be fully controlled. 
The formulation proposed here
on the other hand 
avoids the use of a temperature (while 
nevertheless it does result in 
a Clausius-like formulation when 
appropriate conditions are met, as seen 
above). Additionally, this formulation 
is based on
the degrees of freedom of the system alone, making it 
more
experimentally accessible. 
Furthermore, this definition allows 
to connect negative entropy production 
rates to memory effects, as will be seen 
later in Sec. \ref{sec:EPR_and_Pdiv}.

Finally, integrating Eq.~\eqref{eq:EPR_kinda_Clausius}
we find that the total entropy production over the time interval $[0,t]$
reads \cite{Deffner2011}
\begin{eqnarray}
    \Sigma_S(t) &=&  \int_0^t  ds \; 
    \sigma_S(s)  \nonumber \\
    &=& \Delta S_S(t) + 
    \int_0^t ds \; \Tr \{\dot{\rho}_S (s) 
    \log \rho_S^\star (s) \} \nonumber\\
    &=& D(\rho_S(0) \vert \vert \rho_S^\star (0)) -   D(\rho_S(t) \vert \vert \rho_S^\star (t))  \nonumber\\
    && - \int_0^t ds 
    \Tr \{ \rho_S(s) \partial_s \log \rho_S^\star(s) \}.
\end{eqnarray} 
Note that the last term vanishes when the IFP is time independent. In this
case the total entropy produced in $[0,t]$ reduces to the change of the relative entropy
over this time interval \cite{Deffner2011,Schlögl1980}.
On the basis of the above considerations we will interpret in the following the positivity 
of the entropy production rate,
\begin{equation} \label{gen-second-law}
 \sigma_S(t)\geq 0,
\end{equation}
as a kind of generalized second law in nonequilibrium quantum thermodynamics which
expresses the irreversible character of the dynamics.


\subsection{Relation between positivity of 
entropy production rate 
and the spectrum of the generator}
\label{sec:neg_eigvals}

In this section we study the relation between 
the spectrum of the generator and
the entropy production rate.
We start by imposing some assumptions. First, we suppose again 
that the IFP is a state. Additionally, it 
is reasonable to further assume that it 
belongs to the image of the dynamical map 
$\Phi_t$ (this assumption will be 
satisfied for most undriven open-system scenarios, 
while it may be violated in the 
presence of strong driving) and that 
it does not lie on the boundary of the state space. 
Then, any state of the form 
$\rho_S(t) = \rho_S^\star(t) + \varepsilon X$ 
will be a possible state reached at time $t$ for 
any traceless, Hermitian matrix $X$ and 
sufficiently small $\varepsilon > 0$. 

What can 
we say then about the eigenvalues of the 
generator? 
Let us take a state at time $t$ of the form 
\begin{eqnarray}
    \rho_S(t) = \rho_S^\star (t) + 
    \varepsilon \sigma_\alpha(t),
\end{eqnarray}
where $\sigma_\alpha(t)$ is one of the 
Hermitian and traceless eigenmatrices of the 
generator. Plugging it into Eq. 
\eqref{eq:EPR_generator} we obtain
\begin{eqnarray}\label{eq:EPR_real_eigval}
    && \sigma_S(t) = \\
    &&
    - \varepsilon \lambda_\alpha (t) 
    \Tr \left\{ \sigma_\alpha (t)  \Big[ \log 
    \left( \rho_S^\star (t)+ \varepsilon 
    \sigma_\alpha(t) \right) -
    \log \rho_S^\star (t) \Big] \right\}, \nonumber
\end{eqnarray}
where $\lambda_\alpha(t)$ is the real eigenvalue 
associated to $\sigma_\alpha(t)$.
What we can notice now is that Eq. 
\eqref{eq:EPR_real_eigval} can be rewritten 
as 
\begin{eqnarray}
    \hspace{-0.5cm}\sigma_S (t)=  - \varepsilon \lambda_\alpha (t) 
    \Big[ &D(\rho_S^\star (t) + \varepsilon 
    \sigma_\alpha(t) 
    || \rho_S^\star(t)) &\nonumber\\
    & + D(\rho_S^\star (t) 
    || \rho_S^\star (t) + \varepsilon
    \sigma_\alpha(t)) &\Big],
\end{eqnarray}
where the symmetrized relative entropy appears. 
Since the relative entropy between two states
is always a positive quantity, we can conclude 
that $\lambda_\alpha(t)$ 
is negative if $\sigma_S(t)$ is positive for all
initial states.

A similar but slightly more complicated proof 
follows for the real part of the complex 
eigenvalues.
For simplicity of notation, we will 
skip in the following the dependence on time $t$.
Then, to look at what happens with the complex 
eigenvalues, 
we define the Hermitian
matrices $X_\alpha \equiv \sigma_\alpha + 
\sigma_\alpha^\dagger$ and $Y_\alpha
\equiv i 
(\sigma_\alpha - \sigma_\alpha^\dagger )$,
where $\sigma_\alpha$ is now one of the 
non-Hermitian eigenmatrices of the 
generator, and 
calculate the entropy production rate
for a state $\rho_1 = \rho_S^\star + 
\varepsilon X_\alpha$ and for a state 
$\rho_2 = 
\rho_S^\star + \varepsilon Y_\alpha$. We 
obtain the following expressions 
for the corresponding entropy production rates:
\begin{eqnarray}
    \sigma_1 &=& - \varepsilon\Tr \Big\{ ( \operatorname{Re} 
    \lambda_\alpha X_\alpha + \operatorname{Im} 
    \lambda_\alpha Y_\alpha ) \nonumber \\
    && \hspace{2cm} \Big[ \log(\rho_S^\star 
    + \varepsilon X_\alpha)  -
    \log \rho_S^\star \Big] \Big\}, \\
    \sigma_2  &=& - \varepsilon \Tr \Big\{ ( \operatorname{Re} 
    \lambda_\alpha  Y_\alpha - \operatorname{Im} 
    \lambda_\alpha X_\alpha) \nonumber\\
    && \hspace{2cm} \Big[ \log(\rho_S^\star
    + \varepsilon Y_\alpha)  -
    \log \rho_S^\star \Big] \Big\}.
\end{eqnarray}

If entropy production rate is positive for 
all initial states
(in the sense of $\geq 0$), the 
sum of the two quantities above should also be. 
Putting them together and rewriting in 
terms of relative entropies we find 
\begin{eqnarray}
    &&\hspace{-2.5cm}\sigma_1 + \sigma_2 \\
    = - \varepsilon \Big\{
    \operatorname{Re} \lambda_\alpha 
    &\Big[& D(\rho_S^\star + \varepsilon X_\alpha 
    || \rho_S^\star) + D (\rho_S^\star 
    || \rho_S^\star + \varepsilon X_\alpha)  
    \nonumber\\
    &&+D(\rho_S^\star + \varepsilon Y_\alpha
    || \rho_S^\star) + D (\rho_S^\star |
    | \rho_S^\star + \varepsilon Y_\alpha ) 
    \Big] \nonumber \\
    + \operatorname{Im} 
    \lambda_\alpha &\Big[&D(\rho_S^\star
    || \rho_S^\star + \varepsilon Y_\alpha)
    - D(\rho_S^\star+ \varepsilon Y_\alpha 
    || \rho_S^\star)
    \nonumber\\
    && + D(\rho_S^\star
    + \varepsilon X_\alpha || \rho_S^\star) 
    -D(\rho_S^\star|| \rho_S^\star+
    \varepsilon X_\alpha) 
    \nonumber\\
    &&\hspace{-2.5cm}+ D(\rho_S^\star
    + \varepsilon Y_\alpha || \rho_S^\star
    + \varepsilon X_\alpha) 
     - D(\rho_S^\star
    + \varepsilon X_\alpha || \rho_S^\star
    + \varepsilon Y_\alpha)\Big] 
    \Big\}. \nonumber
\end{eqnarray}
Here the real part of the eigenvalue appears 
again multiplied by the symmetrized relative 
entropy, while the imaginary part is multiplied
by what we could call the antisymmetrized 
relative entropy (in addition to the $\varepsilon$ 
in front of the expression). It can be proven 
(for details see App. 
\ref{app:expansion_relative_entropy}) that 
the symmetrized entropy scales with $\varepsilon^2$, 
while the antisymmetrized, or the asymmetry 
in the relative entropy, is of order 
$\varepsilon^3$ \cite{Audenaert2013}. Thus, for sufficiently small 
$\varepsilon$, only the term with the real part of 
the eigenvalue remains. This term is composed 
of a sum of relative entropies, which are always 
positive, multiplied by $- \text{Re} \lambda_\alpha$. 
Thus, we conclude
\begin{eqnarray}\label{eq:relation_neg_eigvals}
    \sigma(t) \geq 0 \quad \forall \rho_S(0) \quad
    \Rightarrow \quad \text{Re} \,
    \lambda_i(t) \leq 0 \ \forall i,
\end{eqnarray}
which was to be shown. 
The opposite direction of Eq. 
\eqref{eq:relation_neg_eigvals} is, to our 
knowledge, not necessarily true. 

On the other hand, negating the logical relation in Eq. 
\eqref{eq:relation_neg_eigvals} means that, if 
one of the eigenvalues in the spectrum of the 
generator has a positive real part, and 
the map fulfills our assumption that a small 
volume around the IFP always 
belongs to the image 
of the map, then there will exist an 
initial state for which the second law is violated. 
This allows for predictions about the second law 
based on the spectrum of the generator.


\section{Connection between entropy 
production rate and non-Markovianity}
\label{sec:EPR_and_Pdiv}

Next, we will study 
the connection between our proposed
formulation for entropy production rate and 
the presence or lack of memory effects, understood
here as P-divisibility. 
First, we prove that P-divisibility implies 
a positive entropy production rate at all times, 
for all possible initial states. Afterwards we 
show that the converse is not true by means of 
a counterexample which employs the 
phase-covariant model for qubit dynamics. 

\subsection{P-divisibility implies
a positive entropy production rate}
\label{sec:Pdiv_means_second_law}

Following the argumentation in 
\cite{CollaPhD}, let us assume that the
dynamical map $\Phi_t$ is 
P-divisible. This implies that,
if we 
fix the time $t$, the 
superoperator $\mathcal{L}_t$ is the 
generator of a positive semigroup, given by 
the instantaneous map defined in 
Eq.~\eqref{eq:instantaneous_map}.
Additionally, as shown by the 
Brouwer fixed point theorem, 
any positive semigroup has a fixed point which 
is also positive (and thus, a state)
\cite{Davis1976_fe}. 
Thus, entropy production is well 
defined for any P-divisible map. 

Because the IFP in Eq.~\eqref{eq:EPR_Spohn_generalized}
is invariant under the application of the 
instantaneous map in Eq. \eqref{eq:instantaneous_map}, 
we can rewrite entropy production rate 
as 
\begin{eqnarray}\label{eq:EPR_maps}
    \sigma_S (t) = - \left. \frac{d}{d \tau} \right|_{\tau = 0}
    D\left( 
        \Phi_\tau^{(t)}[\rho_S(t)] || \Phi_\tau^{(t)}
    [\rho_S^\star (t)]\right) .
\end{eqnarray}
It was proven in \cite{Muller-Hermes2017-zg}
that the relative entropy is contractive 
under the application of a positive map to both 
of its arguments, hence the derivative in 
Eq. \eqref{eq:EPR_maps} will be negative, 
making the entropy production rate positive. 

Thus, we have proven that
\begin{eqnarray}
    \text{P-divisibility} \Rightarrow \sigma_S(t) 
    \geq 0 \hspace{0.5cm} \forall \rho_S(0).
\end{eqnarray} 

Lastly, we can turn the logical argument 
around and see that any violation of the 
second law, i.e. a negative entropy 
production rate, can be directly linked to 
a lack of P-divisibility and, hence, 
to non-Markovianity of the map.


\subsection{Does a positive entropy production 
rate imply P-divisibility?}
\label{sec:secondlaw_doesntmean_Pdiv}

Given the above connection between 
Markovianity and entropy production rate,
the question arises 
naturally of whether P-divisibility 
can be concluded from an always positive 
entropy production 
rate. The answer is negative, and we show here a 
counterexample for which P-divisibility 
is broken, but this cannot be told 
from the entropy production rate, 
which remains always positive for all initial states. 
We find this in the phase covariant 
model, a paradigmatic model for qubit dynamics
\cite{Filippov2020, Theret2025, Settimo2022}. 
We therefore present first a compact 
expression for the entropy production rate 
for qubits in terms of the Bloch vector, and then 
show the counterexample itself. 


\subsubsection{Entropy production rate for 
qubits}
\label{sec:EPR_qubits}

In the case of qubits, states can be 
described in terms of their Bloch vectors 
$\vec{v}$ such that
\begin{eqnarray}\label{eq:rho_Bloch}
    \rho = \frac{1}{2} \left( 
    \mathbbm{1} + \vec{v} \cdot \vec{\sigma}
    \right),
\end{eqnarray} 
with $\vec{\sigma} = (\sigma_x, \sigma_y, 
\sigma_z)$ the Pauli matrices.
We now derive a formula for the entropy 
production rate in terms of the Bloch vectors of 
the state of the system $\vec{v}(t)$ and 
the IFP $\vec{v}^\star(t)$.

To use the definition for 
entropy production rate found in 
Eq. \eqref{eq:EPR_generator} we need to 
calculate the 
logarithm of the density matrices. Hence, 
it will be useful to
to rewrite the states in exponential form 
$\rho = e^{\vec{r} \cdot \vec{\sigma} }/ Z$, where 
$Z = \Tr \{e^{\vec{r} \cdot \vec{\sigma} }\}$ 
ensures that the trace of the state is 
equal to one. 
Using the relation $e^{i \theta (\vec{n} \cdot 
\vec{\sigma}) } = \mathbbm{1} \cos \theta + 
i (\vec{n} \cdot \vec{\sigma}) \sin \theta$, where 
$\vec{n}$ is a normalized vector, we find 
\begin{eqnarray}
    \frac{e^{\vec{r} \cdot \vec{\sigma}}}{Z} = 
    \frac{1}{2} \left( \mathbbm{1} + 
    \tanh |\vec{r}| 
    \frac{\vec{r}}{|\vec{r}|} \cdot
    \vec{\sigma} \right).
\end{eqnarray}
We want this expression to describe the same 
state as the one in Eq. \eqref{eq:rho_Bloch}, which 
can straightforwardly only happen if the 
vector $\vec{r}$ has the 
same direction as the Bloch vector $\vec{v}$, 
and further the modulus of this vector 
is $|\vec{r}| = \text{arctanh} |\vec{v}|$. 
Thus,
the expressions are equivalent if we choose 
\begin{eqnarray}\label{eq:vec_r}
    \vec{r} \equiv 
    \text{arctanh} |\vec{v}|\frac{\vec{v}}
    {|\vec{v}|}.    
\end{eqnarray}

Looking back at Eq. 
\eqref{eq:EPR_generator} and writing the state 
and the IFP in exponential form, we obtain the 
entropy production rate
\begin{eqnarray}
    \sigma_S(t) = - \Tr \left\{\dot{\rho}_S(t) 
    \left( \log \frac{e^{\vec{r} (t) \cdot 
    \vec{\sigma}}}{Z(t)} - 
    \log \frac{e^{\vec{r^\star(t)} \cdot 
    \vec{\sigma}}}{Z^\star(t)} \right) \right\},
\end{eqnarray}
which after a straightforward calculation
leads to 
\begin{eqnarray}\label{eq:EPR_qubits}
    \sigma_S(t) = - \left[\vec{r}(t)- 
    \vec{r}^\star(t)\right] \cdot 
    \dot{\vec{v}}(t),
\end{eqnarray}
giving us a compact expression for 
the entropy production rate in terms of 
the Bloch vector.


\subsubsection{Counterexample}
\label{sec:counterexample}

We use the phase covariant master 
equation for qubit dynamics, for which 
criteria for P-divisibility are known
\cite{Filippov2020,Theret2025}. 
The master equation
is given by 
\begin{eqnarray}\label{eq:phase_cov_ME}
    \mathcal{L}_t \left[\rho_S(t) \right] &=&
    - i \left[\frac{\omega_r(t)}{2} \sigma_z, 
    \rho_S(t)\right] \\
    &&+
    \gamma_z (t) \left[ \sigma_z \rho_S(t) 
    \sigma_z  - \rho_S(t) \right] \nonumber\\
    &&+
    \gamma_+ (t) \left[ \sigma_+ \rho_S(t) 
    \sigma_- - \frac{1}{2} \left\{ \sigma_-
    \sigma_+, \rho_S(t) \right\} \right] \nonumber\\
    &&+ 
    \gamma_- (t) \left[ \sigma_- \rho_S(t) 
    \sigma_+ - \frac{1}{2} \left\{ \sigma_+
    \sigma_-, \rho_S(t) \right\} \right], \nonumber
\end{eqnarray}
with a Hamiltonian contribution in 
$\sigma_z$ multiplied by a time-dependent 
frequency $\omega_r(t)$, and 
absorption $\gamma_+(t)$, 
emission $\gamma_-(t)$, and pure dephasing 
$\gamma_z(t)$ 
rates with their corresponding 
dissipators. 

From the master equation \eqref{eq:phase_cov_ME} we obtain 
the Heisenberg equations of motion 
for the Bloch vector:
\begin{eqnarray}\label{eq:heis_counterexample}
    && \hspace{-0.4cm}\dot{v}_{x}(t) = - 
    \omega_r(t) v_y(t) - 
    \left( \frac{\gamma_+(t) + 
    \gamma_-(t)}{2} + 2 \gamma_z(t) 
    \right) v_{x}(t),  \nonumber \\
    && \hspace{-0.4cm}\dot{v}_{y}(t) = 
    \omega_r(t) v_x(t) - 
    \left( \frac{\gamma_+(t) + 
    \gamma_-(t)}{2} + 2 \gamma_z(t) 
    \right) v_{y}(t),  \nonumber \\
    && \hspace{-0.4cm} \dot{v}_z(t) = - (\gamma_+(t) 
    + \gamma_-(t) ) v_z(t) + 
    \gamma_+(t) - \gamma_-(t).
\end{eqnarray}
These can be solved by defining the following functions
\begin{eqnarray}
    && \hspace{-0.7cm} \lambda (t) \equiv \exp \left[ - 
    \frac{\Gamma_+(t) + \Gamma_-(t)}{2} -
    2 \Gamma_z(t)
    \right], 
    \label{eq:lambda}\\
    && \hspace{-0.7cm} \lambda_z(t) \equiv \exp \left[ - 
    \Gamma_+ (t) - \Gamma_- (t)\right],  
    \label{eq:lambdaz}\\
    && \hspace{-0.7cm} t_z (t) \equiv \lambda_z(t) \int_0^t 
    ds \left[ \gamma_+(s) - \gamma_-(s) \right] 
    e^{\Gamma_+ (s) + \Gamma_-(s) 
    }, \label{eq:tz}
\end{eqnarray}
where $\Gamma_{\pm(z)}(t) \equiv \int_0^t ds \; 
\gamma_{\pm(z)}(s)$,
and then the evolution of the Bloch vector
is given by 
\begin{eqnarray}
    \label{eq:vxt}
    v_x(t) &=& \lambda(t) \left[v_x(0) 
    \cos \Omega(t) + v_y(0) \sin \Omega(t)\right], \\
    \label{eq:vyt}
    v_y(t) &=& \lambda(t) \left[- v_x(0) 
    \sin \Omega(t) + v_y(0) \cos \Omega (t) \right],\\
    \label{eq:vzt}
    v_z(t) &=& v_z(0) \lambda_z(t) + t_z (t),
\end{eqnarray}
where we have also made use of 
$\Omega(t) \equiv \int_0^t ds \; \omega_r(s)$.

Additionally, the master equation has a 
unique, time-dependent fixed point 
given by 
\begin{eqnarray}\label{eq:vstar}
    \vec{v}^\star(t) = \left(0, 0, 
    \frac{\gamma_+(t) - \gamma_-(t)}
    {\gamma_+(t) + \gamma_-(t)}\right).
\end{eqnarray}
We restrain ourselves to the 
parameter regime in which the 
instantaneous fixed point is a state 
(for which $|\vec{v}^\star(t)| \leq 1$, 
or equivalently 
$\gamma_\pm(t)$ have the same sign at 
each point in time), and thus 
entropy production is well defined. 

Our aim is to find a non-P-divisible
dynamics for which entropy production rate 
is however always positive. 
Necessary and sufficient conditions 
for P-divisibility of 
the map are given by \cite{Filippov2020} 
\begin{eqnarray}\label{eq:Pdiv_conditions}
    \gamma_{\pm} ( t) &\geq& 0, \label{eq:P_div1}\\
    \sqrt{\gamma_+(t) \gamma_-(t)}
    + 2 \gamma_z(t)  &\geq& 0, \label{eq:P_div2}
\end{eqnarray}
which have to be fulfilled at all times. 

On the other hand, in 
order for the entropy production rate to 
remain positive, 
we need all eigenvalues of the map to have 
negative real parts (as seen before in Sec. 
\ref{sec:neg_eigvals}, 
this is a necessary condition). The eigenvalues 
and eigenmatrices (up to 
normalization) 
of the generator are found to be the 
following: 
\begin{eqnarray}
\label{eq:eigsyst_counterexample}
    && \hspace{-0.7cm} \mathcal{L}_t \left[\rho_S^\star
    (t) \right] = 0 , \label{eq:eigvals_model} \\
    && \hspace{-0.7cm} \mathcal{L}_t \left[\sigma_z
    \right] = - (\gamma_+(t) + 
    \gamma_-(t)) \sigma_z, \nonumber \\
    && \hspace{-0.7cm} \mathcal{L}_t \left[\sigma_{+}\right] 
    = \left[ - i \omega_r(t) - \left( \frac{\gamma_+ (t) + 
    \gamma_-(t)}{2} + 2 \gamma_z(t) \right) \right]
    \sigma_{+} , \nonumber \\
    && \hspace{-0.7cm} \mathcal{L}_t \left[\sigma_{-}\right] 
    = \left[ i \omega_r(t) - \left( \frac{\gamma_+ (t) + 
    \gamma_-(t)}{2} + 2 \gamma_z(t) \right) \right]
    \sigma_{-}. \nonumber 
\end{eqnarray}

With all this information we can 
choose the parameters of the master equation.
In order to reduce the number of parameters, we choose $\omega_r(t) = 0$.
Then the requirement of negative 
eigenvalues, together with keeping 
the instantaneous fixed point a state, 
forces us to choose positive values for 
$\gamma_\pm(t)$. Moreover, 
we need 
\begin{eqnarray}
    -\frac{\gamma_+ (t)+ \gamma_-(t)}{4} < 
    \gamma_z(t) < - \frac{\sqrt{\gamma_+(t) 
    \gamma_-(t)}}{2},
\end{eqnarray}
where the left inequality ensures negative 
real parts of the eigenvalues of the generator, 
which has to be fulfilled at all times to 
ensure a positive entropy production rate (see 
Eqs. \eqref{eq:eigvals_model}), 
whilst the right inequality serves to 
violate the P-divisibility criterium in Eq. 
\eqref{eq:P_div2}, and 
has to hold only at some later point in time. 

We choose 
$\gamma_+(t) = 0.2$ and $\gamma_-(t) = 0.8$.
The function for 
$\gamma_z(t)$ we select to be 
\begin{eqnarray}\label{eq:gzt}
    \gamma_z(t) = \begin{cases}
        0 &\text{ for } t\leq 1\\
        - 0.22 (t-1) 
        &\text{ for } 1 <
        t \leq 2\\
        -0.22 
        &\text{ for } t> 2,
    \end{cases}
\end{eqnarray}
starting with a vanishing rate, 
which makes the map initially 
P-divisible (which allows it in turn to be CP) and then 
linearly decreasing until reaching a value 
which violates P-divisibility. In App. 
\ref{app:CPness_map} we show that the map 
chosen is CP. 

\begin{figure}
    \centering
    \includegraphics[width=0.99\linewidth]{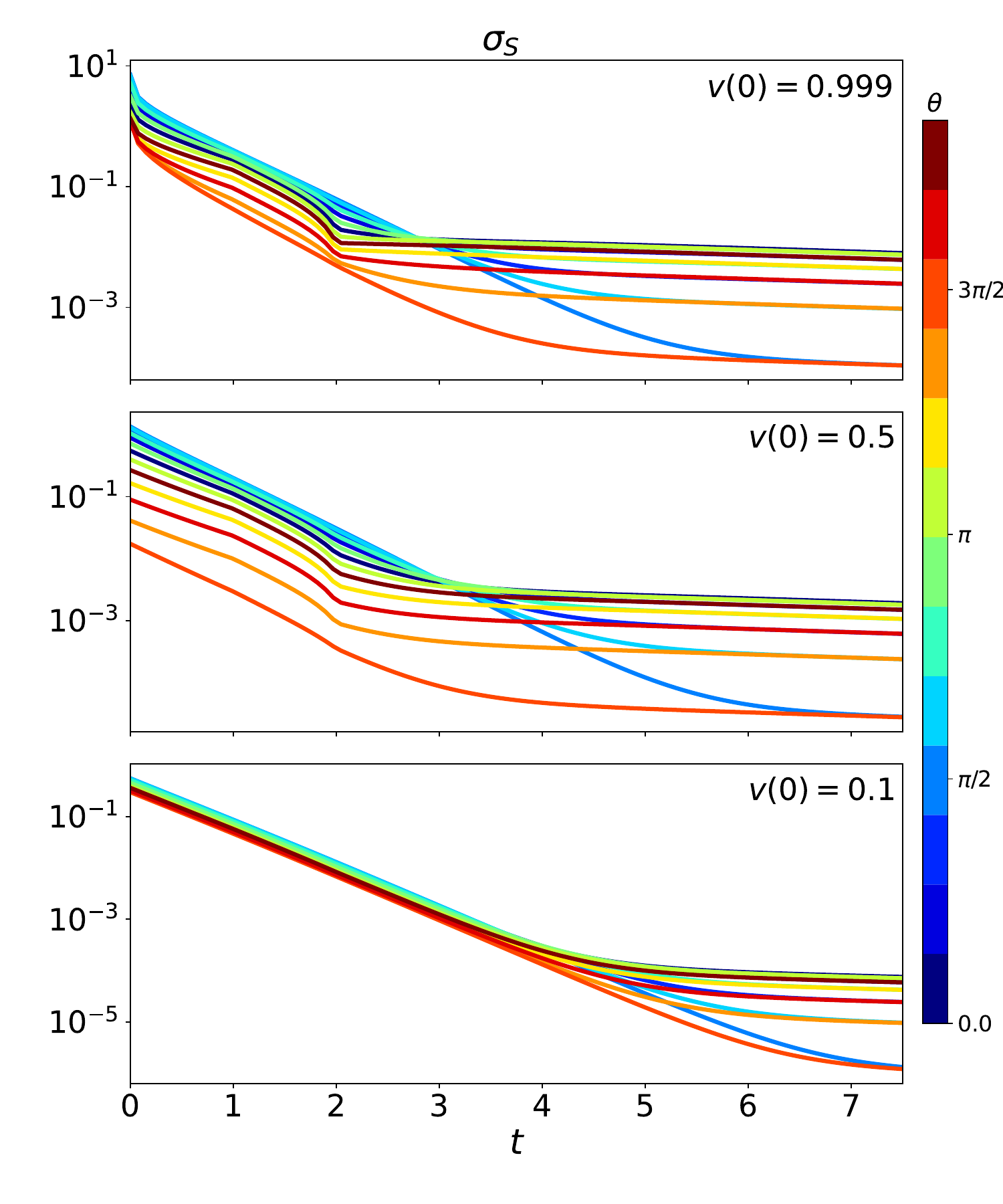}
    \caption{Entropy production rate 
    for the phase-covariant model as 
    a function of time for different 
    initial states with Bloch vector 
    $\vec{v}(0) =v(0) (\cos \theta, 0, 
    \sin \theta)^T$. Each subplot 
    considers a different value of 
    $v(0)$, and each differently colored 
    line represents a different value 
    of $\theta$. The parameters 
    of the master equation are 
    $\gamma_+ = 0.2$, $\gamma_- = 0.8$, and 
    a time dependent $\gamma_z(t)$ which 
    can be found in Eq. \eqref{eq:gzt}.
    The map violates P-divisibility, 
    but this is not witnessed by the 
    entropy production rate, which remains 
    always positive.}
    \label{fig:EPR}
\end{figure}

Using Eq. \eqref{eq:EPR_qubits} we now 
calculate and plot entropy production rate
for different initial states, which 
is shown in Fig. \ref{fig:EPR}. 

Because the dynamics is phase covariant, 
i.e. invariant under rotations around the 
$z$ axis, it suffices to study initial 
states on the $xz$ plane. We therefore take
states of the form 
\begin{eqnarray}\label{eq:ini_states}
    \vec{v}(0) = v(0)(\cos \theta, 
0, \sin \theta)^T,
\end{eqnarray}
where we vary 
both the angle $\theta$ as well as 
the initial radius $v(0)$. 
Additionally, 
since the IFP only has a component in 
the $z$ direction, states 
which are reflected across the $z$ 
axis will result in the same entropy 
production rate, see Eq.~\eqref{eq:EPR_qubits}: If we substitute
$v_x \rightarrow - v_x$, this leads to 
also $r_x \rightarrow - r_x$ and 
$\dot{v}_x \rightarrow - \dot{v}_x$, such 
that the negative signs will cancel each 
other out. Hence, we want to 
choose values $\theta = 
n \frac{2\pi}{N}$, where $N$ is an 
odd total number of values and 
$n$ runs from 0 until $N-1$. This way 
no different initial states are 
reflections of each other across the 
$z$ axis, and 
we  
avoid redundances. 

In Fig. \ref{fig:EPR}
we show the entropy production 
rate as a function of time for different 
initial states of the form given in 
Eq. \eqref{eq:ini_states}. 
We can see that, in all cases, 
entropy production rate remains positive, 
whereas
the choice of $\gamma_z(t)$ 
renders the map non-P-divisible.

We can further study the graph
and we see that
most lines experience an 
abrupt change at the points in 
time when $\gamma_z(t)$ does 
as well. That is, except for the 
lines with $\theta \approx 
\frac{\pi}{2}, \frac{3\pi}{2}$, which 
correspond to initial states 
on the $z$ axis and, as we can see 
from the evolution of the $z$
component of the 
Bloch vector in Eq. \eqref{eq:vzt}, 
the dynamics of these states 
are independent of $\gamma_z(t)$. 
The plot corresponding to $v(0) = 0.1$ 
as well shows this smooth behaviour. 
This seems to indicate that, for states very 
close to the fixed point as well, the $z$ 
contribution dominates over the entropy 
production.
We additionally see that the lines pair 
up for large $t$, in such a way that 
initial states corresponding to 
reflections across the $x$ axis 
end up with the same entropy production 
rate. This happens because, for 
the choice of parameters, the $z$ component 
of the Bloch vector decays towards its 
equilibrium value much faster than 
the other directions. Thus, states
reflected across the $x$ axis end 
up at the same point for large times. 

This numerical evidence supports the 
idea that, for this non-P-divisible map, 
entropy production rate is always positive.
We look nevertheless for an 
analytical proof of this statement.

To start, we want to write down an expression 
for the entropy production in the phase covariant model 
based on the formulation for qubit systems derived 
in  
Eq. \eqref{eq:EPR_qubits}. We recall 
the equations of motion for the 
Bloch vector (see Eqs. 
\eqref{eq:heis_counterexample}) and 
rewrite them as 
\begin{eqnarray}
    \dot{\vec{v}}(t) = \begin{pmatrix}
        - \omega_r(t) v_y(t) + \mu_2(t) v_x(t) \\
        \omega_r(t) v_x(t) + \mu_2(t) v_y(t) \\
        \mu_1(t) \left(v_z(t) - v_z^\star(t)\right)
    \end{pmatrix},
\end{eqnarray}
where we have defined $\mu_{1,2}(t)$ as 
the real part of the eigenvalues of the generator:
\begin{eqnarray}\label{eq:real_part_eigvals}
    \mu_1(t) &=& - \gamma_+(t) - \gamma_-(t) = -1, \\
    \mu_2(t) &=& - \left( \frac{\gamma_+(t) + \gamma_-(t)
    }{2} + 2 \gamma_z(t) \right) = -0.5 - 2 \gamma_z(t). \nonumber
\end{eqnarray}
Using the definition of the 
vector $\vec{r}$ in Eq. \eqref{eq:vec_r},
together with Eq. \eqref{eq:EPR_qubits} 
and with the fixed point in 
Eq. \eqref{eq:vstar}, 
we obtain the entropy production 
\begin{eqnarray}\label{eq:EPR_counterexample_explicit}
    \sigma_S(t) &=& - \mu_2(t) \frac{\text{atanh} 
    |\vec{v}(t)|}{|\vec{v}(t) |} \left( v_x^2(t)
    + v_y^2 (t)\right) \\
    &&- \mu_1(t) (v_z(t) - v_z^\star(t)) \nonumber \\
    &&\hspace{0.5cm} \cdot \left(\frac{\text{atanh} |\vec{v}(t)|}{|\vec{v}(t)|}
    v_z(t) - \frac{\text{atanh} |v_z^\star(t)|}
    {|v_z^\star(t)|}v_z^\star(t)\right). \nonumber
\end{eqnarray}

In the following we will omit the time dependencies 
in order to simplify the notation.
Now we notice that one can rewrite the entropy 
production as a function of three variables: 
$v_z$, $v \equiv |\vec{v}|$ and $\mu_2$:
\begin{eqnarray}\label{eq:EPR_counterexample_simplified}
    \sigma_S &=& 
    - \mu_2 \frac{\text{atanh} 
    v}{v} \left( v^2 - v_z^2 \right) \\
    &&- \mu_1 (v_z - v_z^\star) \cdot 
    \left(\frac{\text{atanh} v}{v}
    v_z - \frac{\text{atanh} |v_z^\star|}
    {|v_z^\star|}v_z^\star\right), \nonumber
\end{eqnarray}
whereas all other parameters (namely, 
$\mu_1$ and $v_z^\star$) are time-independent.
Moreover, the function is restricted to the 
region
\begin{eqnarray}\label{eq:regime}
    -1 \leq v_z \leq 1, 
    \hspace{1cm} |v_z| \leq v \leq 1.
\end{eqnarray}
 
\begin{figure}
    \centering
    \includegraphics[width=0.8\linewidth]{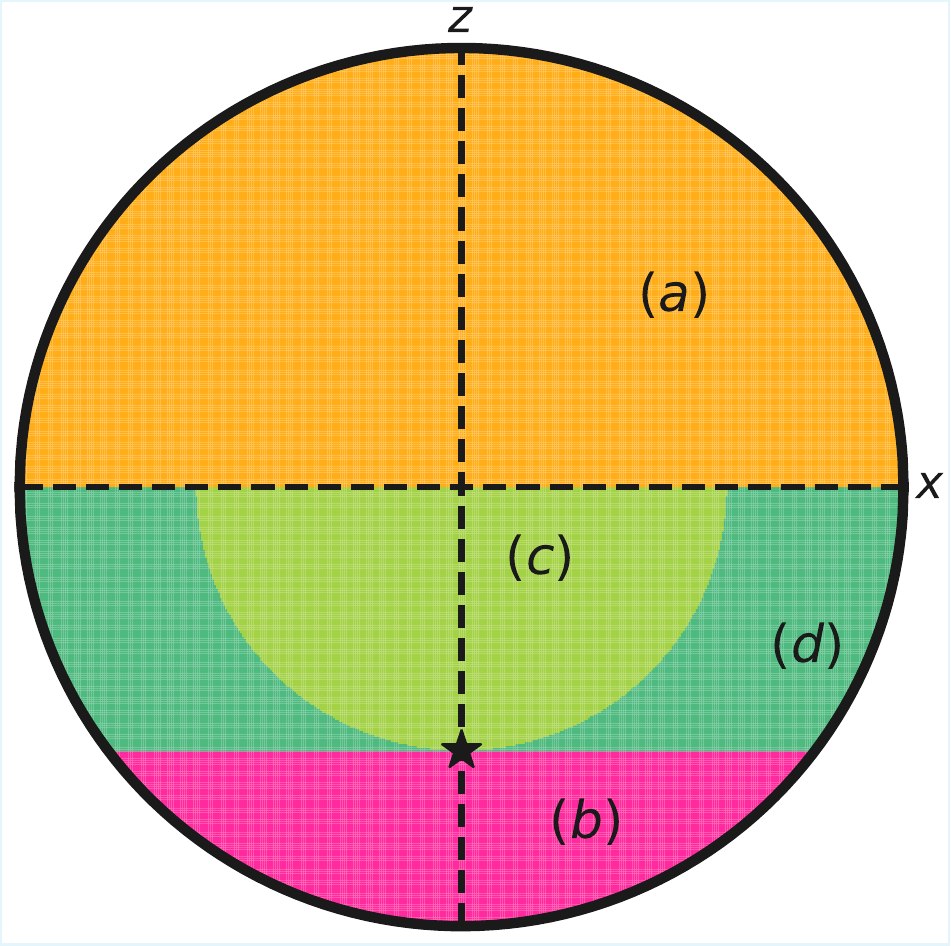}
    \caption{Projection of the Bloch 
    sphere onto the $xz$ plane, where 
    different colors highlight the different 
    regions used in the proof that 
    entropy production rate remains
    always positive.}
    \label{fig:Bloch}
\end{figure}

We divide the proof by distinguishing
four different regions of the 
Bloch sphere, 
and note for the following that for
the parameters chosen $v_z^\star
= -0.6 < 0$ and $\atanh x / x$ is a
positive,
monotonically increasing function
for $x \geq 0$. In Fig. \ref{fig:Bloch} we show 
the different regions
distinguished below. 

(a) $v_z \geq 0$.
Because the real part of the eigenvalues 
is always
negative, the first line of Eq. 
\eqref{eq:EPR_counterexample_simplified} is always positive,
and the second term is where a negative contribution
can appear. Furthermore, 
in this case both factors in the
second part of the entropy 
production rate in Eq.
\eqref{eq:EPR_counterexample_simplified} 
are positive:
\begin{eqnarray}
    v_z-v_z^\star > 0, \hspace{1cm}
    \frac{\atanh v}{v}v_z - 
    \frac{\atanh |v_z^\star|}{|v_z^\star|}
    v_z^\star > 0. \nonumber
\end{eqnarray}
Thus, all terms comprising the entropy production
rate are positive, 
and we can conclude $\sigma_S (t) \geq 0$.
    
(b) $v_z \leq v_z^\star$
This case is similar to the one above, 
because both factors in the 
second line have opposite signs:
\begin{eqnarray}
    v_z-v_z^\star \leq 0, \hspace{1cm}
    \frac{\atanh v}{v}v_z - 
    \frac{\atanh |v_z^\star|}{|v_z^\star|}
    v_z^\star \leq 0. \nonumber
\end{eqnarray}
This means that entropy production rate 
is again always greater or equal to zero.

(c) $0 > v_z > v_z^\star$
and $v \leq |v_z^\star|$.
We first note that
\begin{eqnarray}
    v_z - v_z^\star > 0. \nonumber
\end{eqnarray}
As for the second term, we can argue
\begin{eqnarray}
    &&\frac{\atanh v}{v}v_z - 
    \frac{\atanh |v_z^\star|}{|v_z^\star|}
    v_z^\star \geq \frac{\atanh |v_z^\star|}{|v_z^\star|}
    (v_z - v_z^\star) > 0, \nonumber
\end{eqnarray}
thanks to which the second term in Eq. 
\eqref{eq:EPR_counterexample_simplified}
is also positive, and thus 
$\sigma_S \geq 0$.

(d) $0 > v_z > v_z^\star$ 
and $v > |v_z^\star|$. This part of the 
proof is more involved and we present 
the details of it in 
App. \ref{app:proof_EPR_positive}.

This provides
a counterexample to the hypothesis that 
observation of the second law can 
determine P-divisibility. 
We note however that the parameter regime 
in which we find a counterexample (such 
that the map is CP, but not P-divisible, 
and also all eigenvalues are negative) is 
relatively 
small, signaling that this special 
behaviour is not easy to occur. 

To end our study, we look at a special case 
of the phase-covariant model,
namely when the dynamical map 
becomes a unital map. In this case 
the fixed point is at the center of the 
Bloch sphere, $\vec{v}^\star = \vec{0}$, 
which can only happen if $\gamma_+ (t) = 
\gamma_-(t)$. Then, criteria for 
P-divisibility 
(see Eqs. \eqref{eq:P_div1} and
\eqref{eq:P_div2}) become 
\begin{eqnarray}\label{eq:P-div_unital}
    \gamma_+(t) \geq 0, 
    \hspace{1cm}\gamma_+(t) + 2 \gamma_z(t) \geq 0.
\end{eqnarray}
Additionally, if we look at the 
real part of the eigenvalues in 
Eq. 
\eqref{eq:real_part_eigvals}, we see 
that the conditions for 
these to be negative are equivalent to 
those of P-divisibility in Eq. 
\eqref{eq:P-div_unital}. 
Finally, looking back at the expression 
for entropy production rate found in 
Eq. \eqref{eq:EPR_counterexample_simplified}, 
we see that for a unital map it becomes 
\begin{eqnarray}
    \sigma_S = \frac{\atanh v}{v}
    \left[ - \mu_2
    \left(v_x^2 + v_y^2\right) - 
    \mu_1 v_z^2 \right],
\end{eqnarray}
which is only always positive if 
both $\mu_1, \mu_2 \leq 0$ at all times. 
Thus, for the unital phase-covariant model, 
all three 
properties considered (P-divisibility 
of the map, negativity of the eigenvalues 
and positive entropy production rate) 
become equivalent.

\subsection{Equivalence of non-Markovianity and map entropy production rate}
\label{sec:Camille}

Very recently, during the writing of this manuscript, a similar study on the same
problems has been carried out by Théret \textit{et al.} \cite{Theret2026}. 
The authors of this study compare the 
dynamical behavior of different definitions for the entropy production rate, including our 
definition in Eq. \eqref{eq:EPR_Spohn_generalized}, for the Jaynes-Cummings model, and show 
particularly that two of them coincide: namely the one studied here and the one introduced 
by Esposito \textit{et al.} in \cite{Esposito2010Jan}. 
While this is an interesting result, we note that it does 
not hold in general, as was shown in our recent study of 
pure decoherence models \cite{Picatoste2025_decoherence}.
The authors go on to demonstrate that positivity of entropy
production rate is not equivalent to P-divisibility of the dynamical map. This
is shown numerically employing also a phase covariant model. Coincidentally, it is also 
the model chosen in this work for the 
counterexample which shows agreeing results; 
we provided here additionally an 
analytical proof for the 
fact that entropy production rate remains 
always positive for a non-P-divisible map. 

Finally, Theret \textit{et al.} introduce a function which 
they call the map entropy production, 
defined as 
\begin{eqnarray}
\label{eq:sigma_map}
    \sigma_\text{map} (t) \equiv \min_{\rho_A} 
    \left\{- \Tr \left\{ \mathcal{L}_t [\rho_A] 
    \left[ \log \rho_A - \log \rho_S^\star(t)
    \right]\right\}
    \right\},
\end{eqnarray}
which differs from Eq. \eqref{eq:EPR_generator}
in the fact that the function minimizes 
over all possible input states $\rho_A$ and 
that, for any time $t$, the argument $\rho_A$ is 
allowed to be any positive and trace 1 
matrix, even if this state 
is no longer accessible for the dynamics.
If one recalls, this is precisely the 
kind of consideration we had to employ when proving the positivity of 
entropy production rate in the 
counterexample (see App. \ref{app:proof_EPR_positive}): 
$\sigma_S(t)$ remains positive 
thanks to some states no longer being
available after some time $t$. 
The authors of \cite{Theret2026} consider 
the relation between this quantity and 
P-divisibility of the map at time $t$, 
understood as the existence of $\tau > 0$ such that the 
propagator $\Phi_{t, t^\prime}$  is positive for all
$t \leq t^\prime < t + \tau$. We would also like to study this function in the following. 

To start, when the map is P-divisible at time $t$, 
$\sigma_\text{map}$ can be rewritten 
with a structure reminiscent of 
Eq. \eqref{eq:EPR_Spohn_generalized}: 
\begin{eqnarray}
\label{eq:sigma_map_positive}
    \sigma_\text{map} (t) = \min_{\rho_A }
    \left\{- \left. \frac{d}{d\tau}
    \right|_{\tau = 0} D(
    \Phi_\tau^{(t)} [\rho_A] || \rho_S^\star(t)
    )\right\}.
\end{eqnarray}
Then, as proven in \cite{Theret2026}, the contractivity of the relative entropy under positive maps 
\cite{Muller-Hermes2017-zg} implies 
\begin{eqnarray}\label{eq:P_div_then_sigmamap}
    \text{P-divisibility} \; \Rightarrow \; 
    \sigma_\text{map} \geq 0 .
\end{eqnarray}
Moreover, one can always choose $\rho_A = \rho_S^\star(t)$, which makes 
the quantity to be minimized in Eq.~\eqref{eq:sigma_map} equal to zero, 
which results in $\sigma_\text{map} = 0$ whenever the map is P-divisible. 

However, when P-divisibility is broken then the map entropy production
as defined in Eq. \eqref{eq:sigma_map} is no
longer equivalent to 
Eq. \eqref{eq:sigma_map_positive}, 
since a non-positive instantaneous map $\Phi_\tau^{(t)}$ would map some state $\rho_A$ into a non-positive matrix, and the relative entropy is only a well-defined function when both its arguments are states. Thus, this function loses its interpretation as the change of the relative entropy between a state $\rho_A$ and the instantaneous fixed point under the action of the instantaneous map. 

Nevertheless, this quantity 
is chosen in \cite{Theret2026} because of its connection to memory effects, and the authors show in the case of phase-covariant qubit dynamics that a positive $\sigma_\text{map}(t)$ implies P-divisibility, making the positivity of $\sigma_\text{map}$
a necessary and sufficient condition to determine positivity of the map. 

We want to explore the possibility of generalizing this proof to $d$-level systems. 
Can any lack of P-divisibility be then detected by $\sigma_\text{map}$ for general 
dynamics? To answer this question we first note that according to a theorem by
Kossakowski \cite{Kossakowski1972b,Wissmann2015} a map is P-divisible at time $t$ 
if and only if the generator satisfies the condition
\begin{eqnarray}
    \langle j | \mathcal{L}_t \left( |i \rangle \langle i| \right) | j \rangle \geq0 
    \hspace{0.5cm}\forall i \neq j
\end{eqnarray}
for all orthonormal bases $\{|i\rangle\}$. Thus, if a map is not P-divisible, then 
there is an orthonormal basis $\{|i\rangle\}$ such that
\begin{equation}
 \langle m | \mathcal{L}_t \left( |n \rangle \langle n| \right) | m\rangle < 0
\end{equation}
for a certain pair of states $|n\rangle$, $|m\rangle$. We then take 
a state with spectral decomposition
\begin{eqnarray}
    \rho_A = \sum_{i = 1}^d p_i |i\rangle \langle i|,
\end{eqnarray}
and determine the first term inside the minimization in Eq. \eqref{eq:sigma_map}, namely
\begin{eqnarray} \label{S-dot}
    \dot{S}_S(\rho_A) &=&- \Tr \left\{ \mathcal{L}_t (\rho_A) \log \rho_A\right\} \nonumber \\
    &=&
    - \sum_{i,j = 1}^d p_i \log p_j \langle j | \mathcal{L}_t (|i \rangle \langle i | ) |j\rangle.
\end{eqnarray}
We see that any negative component of the generator, indicating non-Markovianity, appears in the sum. We want thus to choose a state $\rho_A$ in a way that this component prevails.
If $\rho_A$ 
is chosen to be on the boundary of the state space (that is, not full rank: $p_i = 0$ for some $i$), then the corresponding 
logarithmic terms diverge, and will dominate in the sum over $j$. This is precisely the kind of state that is generally not accessible to the dynamics after a certain time, which is what differentiates $\sigma_\text{map}$ from the entropy production rate.

Let us first consider the simplest case of a two-level system. In this case we choose 
$p_n = 1-\varepsilon$ and $p_m = \varepsilon$ with $\varepsilon > 0$ 
such that \eqref{S-dot} becomes
\begin{equation}
\dot{S}_S(\rho_A) = - (\log \varepsilon) (1-\varepsilon) \;  
\langle m | \mathcal{L}_t (|n \rangle \langle n | ) |m\rangle + \ldots,
\end{equation}
where the dots indicate terms which remain finite in the limit $\varepsilon \to 0$.
Thus, we see that $\dot{S}_S(\rho_A)$ approaches $-\infty$ in this limit.
On the other hand, the second term in Eq. \eqref{eq:sigma_map} is always finite (assuming 
the IFP to be full-rank) and, therefore, $\sigma_\text{map}$ becomes $-\infty$.

For a $d$-dimensional system we choose 
$p_n = 1-\varepsilon - \eta$, $p_m = \varepsilon$ and all other $p_i = \eta/(d-2)$ 
(ensuring normalization) with $\varepsilon, \eta> 0$, and obtain
\begin{eqnarray}
\dot{S}_S(\rho_A) &=& - \log \varepsilon \Big[
(1-\varepsilon - \eta) \langle m | \mathcal{L}_t (|n \rangle \langle n | ) |m\rangle
\nonumber \\
&& + \frac{\eta}{d-2} \sum_{i\neq n,m}    
  \langle m | \mathcal{L}_t (|i \rangle \langle i | ) |m\rangle
\Big] + \ldots
\end{eqnarray}
We now choose $\eta$ sufficiently small such that the sign of $\dot{S}_S(\rho_A)$ 
is determined by the first line in the last expression, and then take the limit
$\varepsilon \to 0$. This leads to the same conclusion as before, namely that 
$\dot{S}_S(\rho_A)$ approaches $-\infty$ in this limit and, hence,
$\sigma_\text{map}$ is again equal to $-\infty$.

Summarizing we have shown that any non-P-divisible dynamics will give rise to a negative 
$\sigma_\text{map}$ (or, conversely, that if $\sigma_\text{map} \geq 0$ then there cannot 
be any violation of P-divisibility) which, together with Eq.~\eqref{eq:P_div_then_sigmamap}, leads to the result
\begin{eqnarray}
    \text{P-divisibility} \; \iff \; \sigma_\text{map} 
    \geq0,
\end{eqnarray}
for any finite dimensional system. However, as we have seen $\sigma_\text{map}$
only takes two possible values, namely $\sigma_\text{map}=0$ if the map is P-divisible,
and $\sigma_\text{map}=-\infty$ if the map is not P-divisible. Hence, this quantity
should be interpreted as a binary variable signifying whether or not the dynamics
is P-divisible, rather than as an entropy production rate of the map.

\section{Discussion and conclusions}
\label{sec:conclusions}

In this work we have studied a 
local definition for the entropy production rate of open
quantum systems. This definition 
(given by Eq.~\eqref{eq:EPR_Spohn_generalized})
considers the rate of change of 
the relative entropy between the 
state of the system at each point
in time and an instantaneous fixed point, 
obtained from the time-local 
generator for the evolution of 
the system. It is thus a purely local 
definition, 
based on the degrees 
of freedom of the system alone, and  
therefore 
experimentally accessible for all 
kinds of system-environment couplings 
as long as  
the dynamics of the system can be fully 
kept track of. 

The main restriction to it so far is that 
it can only consider the cases in which 
the IFP is a state. This limits the 
study of e.g. some strong-coupling scenarios, 
and 
the relaxation of this condition is 
left for future work. 
On the other hand, there are two 
main advantages of this definition: 
its connection to memory effects and 
its independence from the first law of 
thermodynamics and from the definition 
of a temperature for the environment, which 
may be non-trivial in the cases where the 
environment is not a thermal bath or, 
even when it is initiated as such, 
when the evolution changes the internal 
state of the bath. 

The connection to memory effects comes 
from the informational nature of the 
formulation: it measures the 
distinguishability of the 
state with respect to the IFP. When 
this distinguishability increases, we 
have shown that it can 
be directly link to the presence of 
memory effects in the dynamics. 
Meanwhile, the choice of the 
relative entropy for the function 
measuring this distinguishability 
is justified by the fact that, when the 
IFP can be written as a Gibbs state of 
the Hamiltonian determining the 
heat exchange, our formulation 
gives rise to the well-known 
Clausius expression, which is recovered 
albeit with some kind of modification. 
The further study of the connection 
between the relative-entropy formulation 
and a Clausius-like expression in more 
general scenarios is 
also left as the subject for future work.  

Given the proposed formulation, 
we have proved the 
following logical chain: 
\begin{eqnarray}
    \text{P-divisibility} \; \Rightarrow  \;
    \sigma_S(t) \geq 0\; \Rightarrow \;
    \text{Re} \, \lambda_i(t) \leq 0,
\end{eqnarray}
where Markovianity of the dynamics is 
understood as P-divisibility of the map,
$\lambda_i(t)$ are the eigenvalues 
of the generator, and $\sigma_S(t) \geq 0$ has 
to hold at all times, for all initial states. 
This provides better understanding of 
the connection between Markovianity, 
the second law of thermodynamics and 
the spectrum of the generator governing 
the dynamics. By contrast, we have also 
shown by means of a counterexample 
that the reverse direction of 
the first statement is not true: 
an always positive entropy production 
rate is not sufficient for a 
dynamics to be P-divisible. 

Moreover, during the writing of this manuscript
a similar study was published by Théret \textit{et al.} \cite{Theret2026}. The authors also study the fixed-point formulation for entropy production rate and additionally propose an extension of the entropy production to the dynamical map, which they show to be equivalent to P-divisibility for phase-covariant qubit dynamics. We here also studied this recent work and extended this connection 
to general $d$-level systems. 

All in all, with this work we hope to clarify some fundamental concepts about entropy 
production for open quantum systems and to pave the way for future research on strong 
coupling quantum thermodynamics.

\acknowledgments
The authors would like to thank Camille L. Latune, Guillaume Théret and Dominique Sugny 
for interesting discussions. 
I.A.P. acknowledges financial support from the 
Studienstiftung des deutschen Volkes. A.C. acknowledges support by the DFG funded 
Research Training Group ``Dynamics of Controlled Atomic and Molecular Systems'' (RTG 
2717).


\appendix


\section{Perturbation expansion of the 
relative entropy}
\label{app:expansion_relative_entropy}

In the following we derive the perturbation 
expansion of the relative entropy between two 
states
\cite{Scandi2025Jul,Audenaert2013,Audenaert_bounds}.
First of all we write the integral expression of 
the logarithm of a matrix
\begin{eqnarray}
    \log A = \int_0^\infty ds \left[
        \left( \mathbbm{1} + s\right)^{-1} - 
        \left( A + s\right)^{-1}
    \right].
\end{eqnarray}
We want then to Taylor expand the function 
$g(\varepsilon) = \log(\rho + \varepsilon X)$ around 
$\varepsilon = 0$, 
where $\rho$ is a full-rank state and $X$ is 
a traceless Hermitian matrix. The zeroth order
term will be simply
$g(0) = \log \rho$. For the first order 
we compute the first derivative of the function 
$g$ with respect to $\varepsilon$, indicated 
by $g^\prime$: 
\begin{eqnarray}
    g^\prime (\varepsilon) = - \int_0^\infty   
    ds \frac{d}{d\varepsilon} (\rho + 
    \varepsilon X + s)^{-1}.
\end{eqnarray}
The derivative of an inverse matrix has to 
be carefully calculated, given that it may not 
commute with the matrix itself. Using the 
identity $\mathbbm{1} = M^{-1}(\varepsilon) 
M(\varepsilon)$, together with the chain rule, 
we find that $(M^{-1}(\varepsilon))^\prime = 
-M^{-1}(\varepsilon) M^\prime (\varepsilon) 
M^{-1}(\varepsilon)$.
Thus, 
\begin{eqnarray}
    g^\prime(0) = \int_0^\infty ds 
    (\rho + s)^{-1} X (\rho + s)^{-1}.
\end{eqnarray}
Higher order terms can be similarly calculated, 
with the second order reading 
\begin{eqnarray}
    \hspace{-0.3cm}g^{\prime \prime}(0) = -2 \int_0^\infty
    ds (\rho+s)^{-1} X (\rho + s)^{-1} X 
    (\rho + s)^{-1}.
\end{eqnarray}
With this we can expand the logarithm term in the 
relative entropy such that 
\begin{eqnarray}
    && \hspace{-0.3cm}D(\rho + \varepsilon X || \rho) \nonumber \\
    && \hspace{-0.3cm}= 
    \Tr \left\{ (\rho + \varepsilon X) 
    \left[ \varepsilon g^\prime(0) + 
    \frac{\varepsilon^2}{2} g^{\prime \prime}
    (0) + \mathcal{O}(\varepsilon^3)
    \right]
    \right\}.
\end{eqnarray}
Thus, the first order term of the relative 
entropy will read 
\begin{eqnarray}
    \varepsilon \Tr \left\{
    X \int_0^\infty (\rho+s)^{-1} \rho 
    (\rho + s)^{-1}
    \right\},
\end{eqnarray}
where we have made use of the cyclic property of 
the trace. The integral is equal to 1, 
since all the terms inside commute, and with 
$X$ being traceless, this term vanishes.

For the second-order contribution to the 
relative entropy we have 
\begin{eqnarray}
    &&\varepsilon^2 \Tr \left\{ 
    \int_0^\infty ds X \left(\mathbbm{1} -   
    (\rho+s)^{-1} \rho \right) (\rho+s)^{-1} X (\rho+s)^{-1}
    \right\} \nonumber \\
    &&= \varepsilon^2 \Tr \left\{
    \int_0^\infty ds \; s X (\rho+s)^{-2} X (\rho+s)^{-1}
    \right\} \nonumber \\
    && = \frac{\varepsilon^2}{2} \Tr \left\{
    X
    \int_0^\infty ds \; (\rho+s)^{-1} X 
    (\rho+s)^{-1} \right\},
\end{eqnarray}
where between the first and second lines 
we have used the identity $\mathbbm{1} - 
(\rho+s)^{-1} \rho = s(\rho+s)^{-1}$ and
the last line comes from integrating by
parts. 

Defining a quantum Fisher scalar product between 
two matrices $A$ and $B$ with respect to a 
reference state $\rho$
\cite{Scandi2025Jul}:
\begin{eqnarray}
    K_{\rho} (A, B) = \Tr \left\{ A 
    \int_0^\infty ds (\rho+s)^{-1} B 
    (\rho+s)^{-1} \right\},
\end{eqnarray}
we can write the relative entropy as
\begin{eqnarray}
    D(\rho+\varepsilon X|| \rho) = 
    \frac{\varepsilon^2}{2} K_\rho(X,X) + 
    \mathcal{O} (\varepsilon^3).
\end{eqnarray}
Here we can see that the second order 
contribution to the relative entropy is 
proportional to the Fisher square norm of 
the traceless matrix $X$ with respect to 
the reference state $\rho$, which makes 
the relative entropy
symmetric in its arguments up to second order. 
The asymetry can be at maximum 
third order in the perturbation parameter.


\section{Complete positivity of the map}
\label{app:CPness_map}

\begin{figure}
    \centering
    \includegraphics[width=0.99\linewidth]{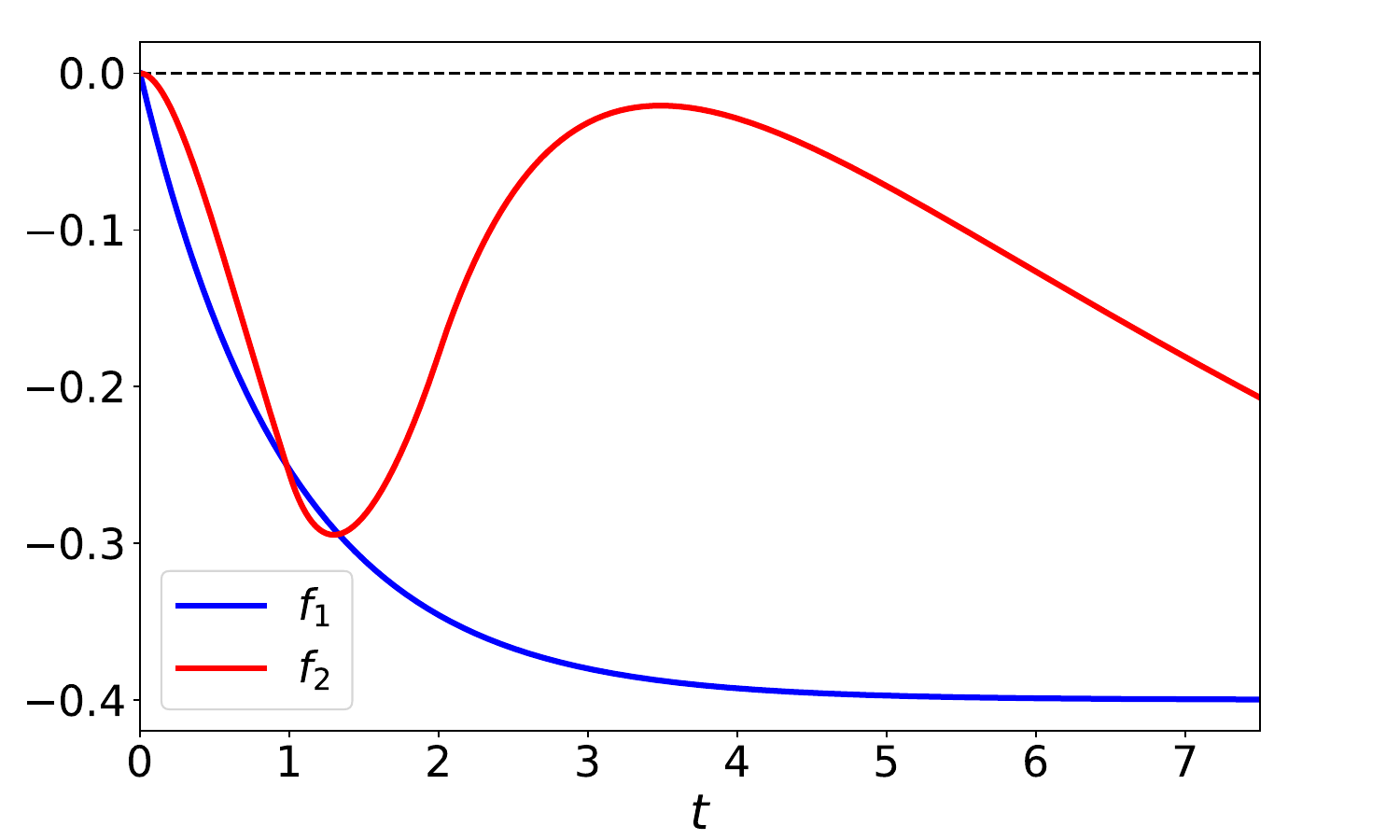}
    \caption{Functions from Eq. \eqref{eq:conds_CP}
    which determine complete positivity of the map. 
    The fact that both $f_1(t)$ and $f_2(t)$ remain
    smaller or equal to zero confirms that 
    the map studied is CP.}
    \label{fig:CP_conds}
\end{figure}

The conditions for the full map 
to be completely positive are given by 
\cite{Filippov2020}
\begin{eqnarray}
    \label{eq:conds_CP}
    f_1(t) &\equiv& |\lambda_z(t) | + |t_z(t) | - 
    1 \leq 0 ,\\
    f_2(t) &\equiv& 4 \lambda^2(t) + t_z^2(t) - 
    \left[ 1+ \lambda_z(t) \right]^2 \leq 0.
\end{eqnarray}
In Fig. \ref{fig:CP_conds} we plot functions
$f_{1,2}(t)$ as a function of time for the 
parameters chosen in the counterexample 
(see Fig. \ref{fig:EPR} in Sec. \ref{sec:counterexample}), showing 
that the conditions are fulfilled and 
the map 
studied is CP. 

\section{Proof of $\sigma_S(t) \geq0$ for 
the counterexample}
\label{app:proof_EPR_positive}

We show here a proof of the 
positivity of entropy production rate 
in region (d) of the Bloch 
sphere (see Fig. \ref{fig:Bloch})
for the counterexample shown in Sec. 
\ref{sec:counterexample}.

Region (d) is defined by 
$0 > v_z > v_z^\star$ 
and $v > |v_z^\star|$. This last proof 
becomes more convoluted, and we 
have to
divide it into three 
time intervals: first $t\leq 1$, then
$1 < t \leq 1.5$ and finally $t > 1.5$. 

We start by looking at 
the entropy production rate for 
$t \leq 1$. For this time 
interval $\gamma_z(t) \equiv 0$ and 
leads to a value of 
$ \mu_2 = -0.5 \equiv \mu_{2, \text{min}}$.
For this value we rewrite our expression 
as 
\begin{eqnarray}
    \sigma_S &=& 
    \frac{\text{atanh} 
    v}{v} \left[ - \mu_{2, \text{min}} \left( 
    v^2 - v_z^2 \right) 
    - \mu_1 (v_z - v_z^\star) v_z \right]
    \nonumber\\
    &&+  \frac{\text{atanh} |v_z^\star|}
    {|v_z^\star|} 
    \mu_1 (v_z - v_z^\star)v_z^\star. 
\end{eqnarray}
The second line in this expression 
is always positive, as well as the 
prefactor multiplying the first line, 
so we have to look at the behaviour of 
the parabola 
\begin{eqnarray}
    &&(\mu_{2,\text{min}} - 
    \mu_1) v_z^2 + \mu_1 v_z^\star v_z 
    - \mu_{2,\text{min}} v^2 \nonumber\\
    &&\geq (\mu_{2,\text{min}} - 
    \mu_1) v_z^2 + \mu_1 v_z^\star v_z 
    - \mu_{2,\text{min}} |v_z^\star|^2 
    \nonumber,
\end{eqnarray}
where in the second line we have 
made use of conditions that 
describe region (d). 
Plugging in the values for 
the parameters, it can be seen 
that the second line becomes $0.5 v_z^2
+0.6 v_z + 0.18$, describing
an upwards-open parabola 
with a minimum value of $0$ at 
$v_z = -0.6$. Thus, this function 
is always non-negative and, together 
with the other terms, 
we can prove that
$\sigma_S \geq 0$ for all possible 
states in region (d) for as long as 
$\mu_2 = \mu_{2,\text{min}}$ (that is, 
for $t \leq 1$). 
    
For $t>1$ the parameter 
$\mu_2$ changes in time 
along with $\gamma_z(t)$ (recall 
Eq. \eqref{eq:gzt}), and when it does 
the parabola no longer remains positive. 
However, when $\mu_2$ differs 
from its minimum value (at $t>1$), 
all states have evolved, and not 
all combinations of $v$ and $v_z$ 
represent accessible states. 
In particular, there is a 
maximum value that $v$ can take
at each point in time. 

To find this maximum distance from 
the center of the Bloch sphere we look 
at the evolution of the Bloch 
vector presented in Eqs. \eqref{eq:vxt}-
\eqref{eq:vzt}, and calculate
\begin{eqnarray}
    \hspace{-0.2cm}v^2(t) = \left[v^2(0) - v_z^2(0)
    \right] \lambda^2(t) + \left[
    t_z(t) + v_z(0) \lambda_z(t)
    \right]^2. 
\end{eqnarray}
Using the fact that $v(0) \leq 1$, 
and rearranging, we obtain 
\begin{eqnarray}\label{eq:rmax_ineq}
    v^2(t) \leq &&\left[
    \lambda_z^2(t) - \lambda^2(t)
    \right] v_z^2(0) \nonumber \\
    &&+ 2 \lambda_z(t) 
    t_z(t) v_z(0) + t_z^2 (t) 
    + \lambda^2(t).
\end{eqnarray}
Here on the right-hand-side we again have 
another parabolic function of 
$v_z(0)$. Introducing 
$x = v_z(0)$ we  
rewrite it as 
\begin{eqnarray}\label{eq:parabola_rmax}
    P(x) \equiv&& \left[
    \lambda_z^2(t) - \lambda^2(t)
    \right] x^2 \nonumber \\
    &&+ 2 \lambda_z(t) 
    t_z(t) x + t_z^2 (t) 
    + \lambda^2(t).
\end{eqnarray} 

\begin{figure}
    \centering
    \includegraphics[width=0.99\linewidth]{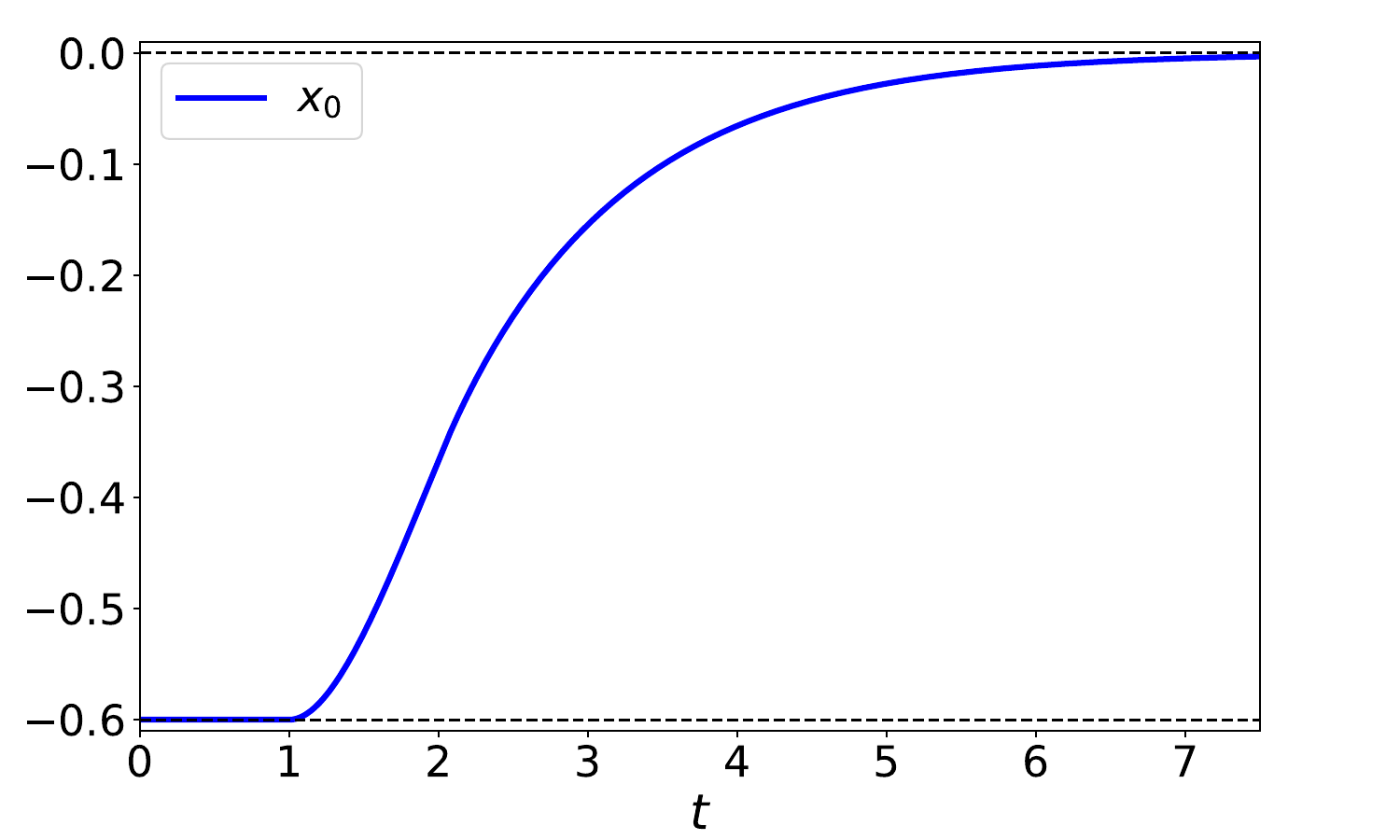}
    \caption{$x_0$ function (blue solid line)
    defined in Eq. \eqref{eq:x0} as 
    a function of time, along with 
    the horizontal black dashed lines
    marking the boundary values $0$ and 
    $-0.6$. }
    \label{fig:x0}
\end{figure}

In order to find its maximum, we 
need to study the coefficients appearing in 
it. 
From Eqs. \eqref{eq:lambda}-\eqref{eq:tz}, 
together with \eqref{eq:gzt}, we find 
the explicit expressions for the 
parameters chosen for the counterexample:
\begin{eqnarray}\label{eq:lambda_explicit}
    \lambda(t) = \begin{cases}
        e^{-0.5 t} &\text{ for } t\leq 1\\
        e^{0.22 - 0.94 t + 0.22 t^2} 
        &\text{ for } 1 <
        t \leq 2\\
        e^{-0.66 - 0.06 t}
        &\text{ for } t> 2,
    \end{cases} 
\end{eqnarray}
\begin{eqnarray}\label{eq:lambdaz_explicit}
    \lambda_z(t) = e^{-t},
\end{eqnarray}
\begin{eqnarray}\label{eq:tz_explicit}
    t_z(t) = -0.6 (1-e^{-t}).
\end{eqnarray}
Based on these equations it is easy to see 
that 
$\lambda_z(t) < \lambda(t)$, such 
that the function in Eq. 
\eqref{eq:parabola_rmax} is a 
downward open parabola. 
The point at which 
its first derivative is zero will 
therefore constitute its maximum, and 
this point is found at 
\begin{eqnarray}\label{eq:x0}
    x_0(t) = \frac{\lambda_z(t)t_z(t)}
    {\lambda^2(t) - \lambda_z^2(t)},
\end{eqnarray}
which can be checked (using 
Eqs. \eqref{eq:lambda_explicit}- 
\eqref{eq:tz_explicit}) to always belong to 
the interval $[-0.6,0]$, as is 
shown in Fig. \ref{fig:x0}.  
At this value the parabola in 
Eq. \eqref{eq:parabola_rmax}
reaches a maximum value of 
\begin{eqnarray}\label{eq:P0}
    P_0 (t) = 
    \frac{\lambda_z^2(t) t_z^2(t)}
    {\lambda^2(t) - \lambda_z^2(t)} 
    + t_z^2(t) + \lambda^2(t).
\end{eqnarray}
Since this is the maximum value that 
the right-hand-side of Eq. 
\eqref{eq:rmax_ineq} can take at 
each point in time, we can conclude 
that $v^2(t) \leq P_0 (t)$, and 
therefore $v(t) \leq \sqrt{P_0(t)}$. 

\begin{figure}
    \centering
    \includegraphics[width=0.99\linewidth]{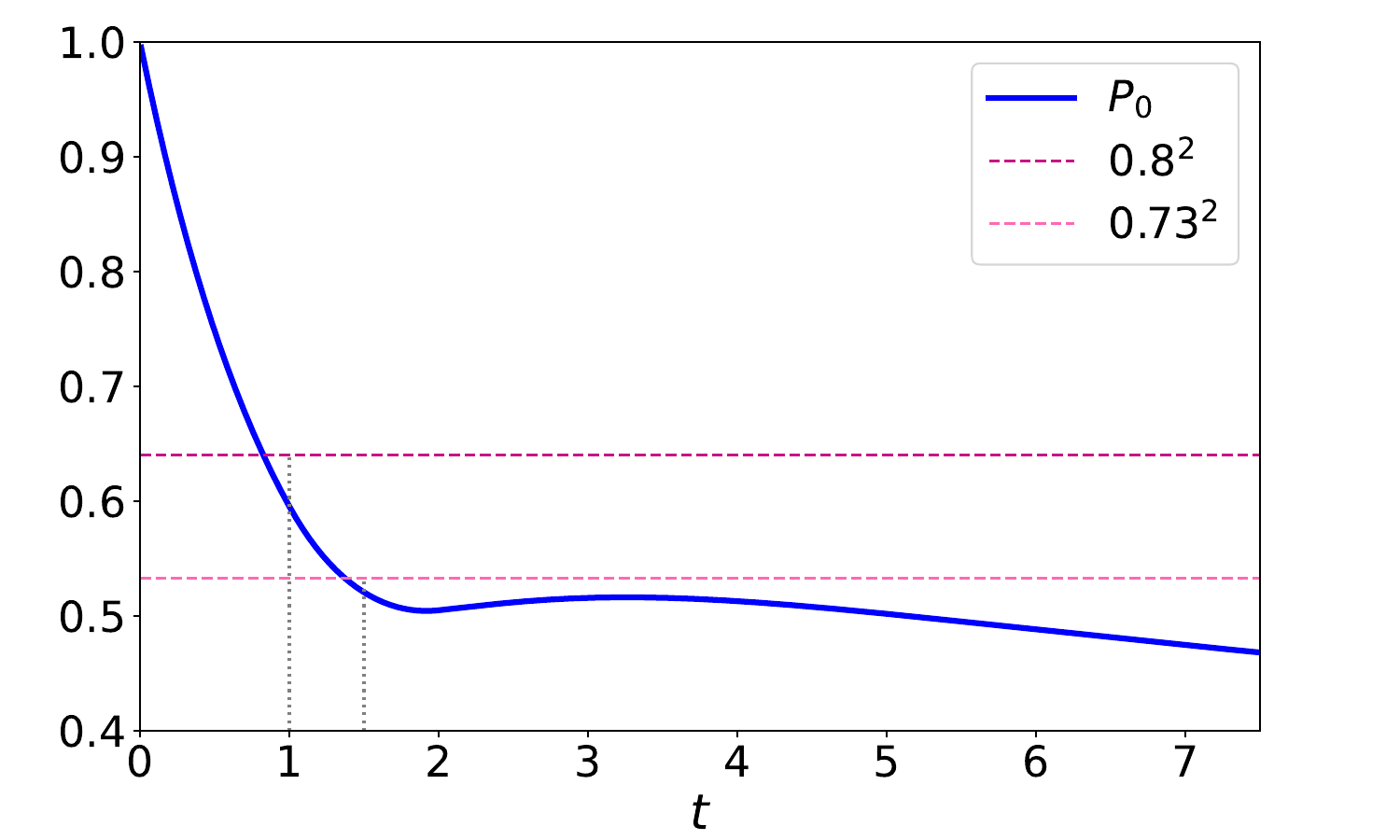}
    \caption{$P_0$ function (blue solid line)
    defined in Eq. \eqref{eq:P0} as 
    a function of time, along with 
    the horizontal dashed lines
    marking the values $0.8^2$ and 
    $0.73^2$. The vertical dotted lines 
    mark the times $t = 1$ and 
    $t = 1.5$.
    Since $v^2(t) \geq P_0(t)$,
    we can see that for $t \geq 1$, 
    the inequality $v(t) 
    \leq 0.8$ holds, and for $t > 1.5$, 
    so does $v(t) 
    \leq 0.73$.}
    \label{fig:P0}
\end{figure}

In Fig. \ref{fig:P0} 
we plot $P_0$ as 
a function of time. The aim is to find 
a value that can bound  
$v$ for the time intervals 
specified above. In particular, we see
that, for $t > 1$, it always holds that 
$P_0 < 0.8^2$ (such that 
$v < 0.8$), and for 
$t > 1.5$ it is true that 
$P_0 < 0.73^2$ (so 
$v < 0.73$). These specific values were chosen for 
convenience, and we will denote by 
$v_\text{max}$ the value for the 
corresponding time interval. 

With this $v_\text{max}$ 
we turn back to our expression for en-
tropy production rate in Eq. 
\eqref{eq:EPR_counterexample_simplified}. 
We start by noticing that, since
the first line in Eq. 
\eqref{eq:EPR_counterexample_simplified} 
is positive, we can bound the
expression through
\begin{eqnarray}
    \sigma_S &\geq& 
    - \mu_{2,\text{max}} \frac{\text{atanh} 
    v}{v} \left( v^2 - v_z^2 \right) \\
    &&- \mu_1 (v_z - v_z^\star) \cdot 
    \left(\frac{\text{atanh} v}{v}
    v_z - \frac{\text{atanh} |v_z^\star|}
    {|v_z^\star|}v_z^\star\right), \nonumber
\end{eqnarray}
where now $\mu_{2,\text{max}}$ is the 
maximum value (i.e., $- \mu_2$ is minimum)
that $\mu_2$ can attain in the time 
interval considered. This value corresponds 
to $\mu_{2,\text{max}} = -0.28$ for $1<t\leq 1.5$
and $\mu_{2,\text{max}} = -0.06$ for 
$t > 1.5$
(see Eqs. \eqref{eq:gzt} and \eqref{eq:real_part_eigvals}). 
Using the
finding that $v \leq v_\text{max}$, 
we further bound the expression
by
\begin{eqnarray}
    \sigma_S &\geq& 
    - \mu_{2,\text{max}} \frac{\text{atanh} 
    v}{v} \left( v^2 - v_z^2 \right) \\
    &&- \mu_1 (v_z - v_z^\star) \cdot 
    \left(\frac{\text{atanh} v_\text{max}}
    {v_\text{max}}
    v_z - \frac{\text{atanh} |v_z^\star|}
    {|v_z^\star|}v_z^\star\right), \nonumber
\end{eqnarray}
(since $-\mu_1$ is positive, as well as 
$v_z - v_z^\star$, but $v_z$ itself is 
negative). 
Next, because of the conditions 
defining region (d) we can bound 
$v^2 > (v_z^\star)^2$ and 
$\atanh v / v > \atanh |v_z^\star| / |v_z^\star|$: 
\begin{eqnarray}
    \sigma_S &\geq& 
    - \mu_{2,\text{max}} \frac{\text{atanh} 
    |v_z^\star|}{|v_z^\star|} 
    \left( (v_z^\star)^2 - v_z^2 \right) \\
    &&- \mu_1 (v_z - v_z^\star) \cdot 
    \left(\frac{\text{atanh} v_\text{max}}
    {v_\text{max}}
    v_z - \frac{\text{atanh} |v_z^\star|}
    {|v_z^\star|}v_z^\star\right), \nonumber
\end{eqnarray}
and rearranging the above expression 
leads to 
\begin{eqnarray}\label{eq:final_bound_d}
    \sigma_S &\geq&
    (v_z - v_z^\star) \times \nonumber \\
    &\times& \left[ v_z \left(
    - \mu_1 \frac{\atanh v_\text{max}}
    {v_\text{max}} + \mu_{2,\text{max}}
    \frac{\atanh |v_z^\star|}{|v_z^\star|}
    \right)\right. \nonumber \\
    && \;\;\; + \left. 
    \frac{\atanh |v_z^\star|}{|v_z^\star|}
    (\mu_{2, \text{max}} + \mu_1) v_z^\star
    \right],
\end{eqnarray}
where the prefactor multiplying the 
expression on the right-hand side is 
always positive in region (d).

Now let us look explicitly at the 
two time intervals left to study, and 
plug in explicit values for all parameters
into Eq. \eqref{eq:final_bound_d}
(with the numbers in 
the functions for $\sigma_S$  
approximated to two decimal places): 
\begin{eqnarray}
    \bullet \hspace{0.5cm} &&1 < t \leq 1.5: \hspace{0.5cm} \begin{rcases}
        \begin{dcases}
        \mu_{2,\text{max}} = -0.28 \\
        v_\text{max} = 0.8
        \end{dcases}
    \end{rcases} \nonumber\\
    &&\Rightarrow
    \sigma_S \geq (v_z-v_z^\star) \left[
    1.05 v_z + 0.88 \right] \geq 0, \nonumber \\
    \bullet \hspace{0.5cm} &&t > 1.5: \hspace{0.5cm} \begin{rcases}
        \begin{dcases}
        \mu_{2,\text{max}} = -0.06 \\
        v_\text{max} = 0.73
        \end{dcases}
    \end{rcases} \nonumber\\
    &&\Rightarrow
    \sigma_S \geq (v_z-v_z^\star) \left[
    1.21 v_z + 0.73 \right] \geq 0, \nonumber
\end{eqnarray}
which are both positive because 
we are considering
the interval $v_z \in (-0.6, 0)$.  
This concludes our proof for the positivity 
of the entropy production rate 
for the counterexample in Sec. \ref{sec:counterexample}.

\bibliography{biblio}
\end{document}